\documentclass[aps,pre,superscriptaddress,citeautoscript, amsmath,amssymb,twocolumn]{revtex4-1}	
\usepackage{graphicx}
\usepackage{amsmath}
\usepackage{hyperref}
\usepackage{color}
\usepackage{bigints}
\usepackage{mathrsfs}
\usepackage{algorithm}
\usepackage{algpseudocode}
\definecolor{darkblue}{RGB}{8,81,156}
\hypersetup{colorlinks,breaklinks,
            linkcolor=darkblue,urlcolor=darkblue,
           anchorcolor=darkblue,citecolor=darkblue}

\date{\today}

\allowdisplaybreaks

    \definecolor{dark-purple}{RGB}{118,42,131}
    \definecolor{dark-green}{RGB}{27,120,55}
    \definecolor{light-purple}{RGB}{231,212,232}
    \definecolor{LIGHT-PURPLE}{RGB}{194,165,207}
    \definecolor{light-green}{RGB}{168,216,183}
    \definecolor{gray}{RGB}{186,186,186}
    \definecolor{super-dark-green}{RGB}{0,69,41}
    \definecolor{super-dark-purple}{RGB}{63,0,125}
    \definecolor{super-dark-blue}{RGB}{8,48,107}
    \definecolor{super-dark-red}{RGB}{165,0,38}
    
    \definecolor{super-dark-purple}{RGB}{64,0,75}
    \definecolor{super-dark-green}{RGB}{0,68,27}

\usepackage{tabularx,booktabs} 

\usepackage{array}
\usepackage{longtable}
\newcolumntype{L}[1]{>{\raggedright\let\newline\\\arraybackslash\hspace{0pt}}p{#1}}
\newcolumntype{C}[1]{>{\centering\let\newline\\\arraybackslash\hspace{0pt}}m{#1}}
\newcolumntype{R}[1]{>{\raggedleft\let\newline\\\arraybackslash\hspace{0pt}}m{#1}}

\setcitestyle{super}

\usepackage{atveryend}
\makeatletter
\let\origcitation\citation
\AtEndDocument{\def\mycites{\@gobble}%
  \def\citation#1{\g@addto@macro\mycites{,#1}\origcitation{#1}}}
\AtVeryEndDocument{\typeout{***^^JCited keys: \mycites^^J***}}
\makeatother

\makeatletter
\let\origcitation\citation
\AtEndDocument{\def\mycites{}%
  \def\citation#1{\g@addto@macro\mycites{#1^^J,}\origcitation{#1}}}
\AtVeryEndDocument{\newwrite\citeout\immediate\openout\citeout=\jobname.cit
  \immediate\write\citeout{\mycites}\immediate\closeout\citeout}
\makeatother

\begin{document}

\title{Induced Charge Anisotropy: a Hidden Variable Affecting Ion Transport through Membranes}

\author{Hessam Malmir}
\email{hessam.malmir@yale.edu}
\affiliation{Department of Chemical and Environmental Engineering, Yale University, New Haven, CT  06520}

\author{Razi Epsztein}
\email{Razi.Epsztein@yale.edu}
\affiliation{Department of Chemical and Environmental Engineering, Yale University, New Haven, CT  06520}

\author{Menachem Elimelech}
\email{enachem.elimelech@yale.edu}
\affiliation{Department of Chemical and Environmental Engineering, Yale University, New Haven, CT  06520}

\author{Amir Haji-Akbari}
\email{amir.hajiakbaribalou@yale.edu}
\affiliation{Department of Chemical and Environmental Engineering, Yale University, New Haven, CT  06520}

\begin{abstract}
The ability of semipermeable membranes to selectively impede the transport of undesirable solutes is key to many applications. Yet, obtaining a systematic understanding of how membrane structure affects selectivity remains elusive due to the insufficient spatiotemporal resolution of existing experimental techniques, and the inaccessibility of relevant solute transport timescales to conventional molecular simulations. Here, we utilize jumpy forward-flux sampling to probe the transport of sodium and chloride ions through a graphitic membrane with sub-nm pores. We find chlorides to traverse the pore at rates over two orders of magnitude faster than sodiums. We also identify two major impediments to the transport of both ion types. In addition to the partial dehydration of the leading ion, its traversal induces charge anisotropy at its rear, which exerts a net restraining force on the ion. Charge anisotropy is therefore a crucial hidden variable controlling the kinetics of ion transport through nanopores.
\end{abstract}

\maketitle

\section{Introduction}
The ability to control ion and solute transport through membranes is central to many processes in chemistry, biology and materials science, such as water desalination~\cite{ShannonNature2008, ElimelechScience2011}, chemical separation of gases~\cite{KorosJMembSci1993, HutchingsNanoscaleHoriz2019}, ions~\cite{ChengJMembSci2018}, organic solvents~\cite{MarchettiChemRev2014}, and viruses~\cite{YangAdvMater2006}, and \emph{in vivo} transport of ions, pharmaceuticals and nutrients through biological membranes~\cite{GurtovekoJACS2005} and channel proteins~\cite{KumarPNAS2007}. Most such applications rely on the semipermeability of the underlying membrane, i.e.,~its ability to preferentially allow  for the passage of some molecules and/or ions while excluding the majority of other components. Consequently, the need to improve solvent permeability and solvent-solute selectivity of membranes  has been extensively addressed in recent years~\cite{HoltScience2006, GeiseJMembraneSci2011, NairScience2012, LiuACSNano2014, WerberNatRevMater2016, SakiyamaNatNanotechnol2016, WalkerACSNano2017, LivelyNatMater2017, RaziEnvSciTech2018, RaziJMembrSci2018, BooEnvirSciTech2018}. 
The major obstacle to enhancing solute-solute selectivity is the considerable gap in our understanding of the molecular-level features that control selectivity. This is primarily due to insufficient spatiotemporal resolutions of most experimental techniques in probing the molecular mechanism of solvent and solute transport through nanopores. In principle, it is generally understood that the selectivity of a nanoporous membrane for a specific solute is mainly dictated by steric~\cite{VanderbrugenWaterRes2004}, charge-exclusion~\cite{DeenAICheJ1987} and solvation~\cite{CorryJPhysChemB2008, SongJPhysChemB2009, RichardsPCCP2012, RichardsSmall2012, SahuNanoscale2017, EpszteinJMembrSci2019} effects. More recent investigations, however, point to a more complex picture, and underlines the importance of other more subtle factors such as polarizability effects~\cite{GrosjeanNatComm2019} and mechanical properties of the membrane~\cite{MarbachNatPhys2018}. Consequently, selectivity is affected by a complex interplay of a confluence of factors, and cannot be determined by single measures, such as ion size, charge density or hydration energy, as demonstrated in recent experimental studies~\cite{RaziJMembrSci2018}. Understanding how membrane structure affects its selectivity for different types of solutes is therefore of pressing importance. 

In recent years, there has been an increased interest in using molecular simulations to study solvent and solute transport through nanoporous membranes, as molecular simulations have been utilized for computing properties such as solvent permeability~\cite{CohenNanoLett2012, HeiranianNatComm2015, CohenNanoLett2016, ChenACSApplMaterInt2017, DahanayakaPCCP2017}, free energy barriers~\cite{RichardsPCCP2012, RichardsSmall2012, KonathamLangmuir2013, CohenNanoLett2016, SahuNanoLett2017, SahuNanoscale2017}, and solute rejection rates~\cite{CohenNanoLett2012, HeiranianNatComm2015, ChenACSApplMaterInt2017} across numerous well-defined nanoporous membranes.  However, these studies either employ conventional techniques such as molecular dynamics (MD)~\cite{AlderMDJCP1959}, which provide unbiased kinetic and mechanistic information but cannot efficiently probe long solute passage timescales, or utilize techniques such as umbrella sampling~\cite{TorrieJCompPhys1977} that are based on applying biasing potentials along pre-specified reaction coordinates, but provide no direct information about kinetics. Therefore, such traditional techniques are  inadequate for comprehensively investigating the structure-selectivity relationship in ultra-selective membranes due to their limited range of accessible timescales or their inability to probe the passage kinetics altogether.

Here, we apply non-equilibrium MD simulations and jumpy forward-flux sampling (jFFS)~\cite{HajiAkbariJChemPhys2018} to investigate the pressure-driven transport of sodium (Na$^+$) and chloride (Cl$^-$) ions across multilayer nanoporous graphitic membranes. Nanoporous graphene has been shown in numerous studies~\cite{SintJACS2008, CohenNanoLett2012, KonathamLangmuir2013, CohenNanoLett2016, Tunuguntla2017} to be a potential next-generation desalination membrane. 
  Using an advanced sampling technique such as FFS, which has been successfully utilized for studying rare events, such as crystal nucleation~\cite{ValerianiJChemPhys2005, LiNatMater2009, LiPCCP2011, ThaparPRL2014, HajiAkbariPCCP2014, HajiAkbariPNAS2015, GianettiPCCP2016,  HajiAkbariPNAS2017, JiangJChemPhys2018}, hydrophobic evaporation~\cite{SumitPNAS2012, AltabetPNAS2017}, and protein folding~\cite{BorreroJCP2006}, enables us to precisely and efficiently compute arbitrarily long mean passage times, and to obtain a statistically representative picture of the ion transport mechanism. In the case of the graphitic nanoporous membrane considered in this work, 
we find the first ion to traverse the pore to always be a chloride, with mean passage times of several microseconds. This corresponds to a solute passage ratio of one chloride per 10,000 solvent molecules. Sodium ions, however, traverse the pore over passage times close to a millisecond. This wide separation of timescales between sodium and chloride transport results in the emergence of a reversal potential across the membrane.  We also demonstrate that for both ion types, the kinetics of ion transport is governed by partial dehydration and the reorganization of the hydration shell, as well as the emergence of charge anisotropy in the salty feed during the transport process. This induced charge anisotropy is a previously overlooked hidden variable that strongly impacts the kinetics and mechanism of ion transport.

\section{Methods}

\begin{figure*}[ht]
    \centering
    \includegraphics[width=.7\textwidth]{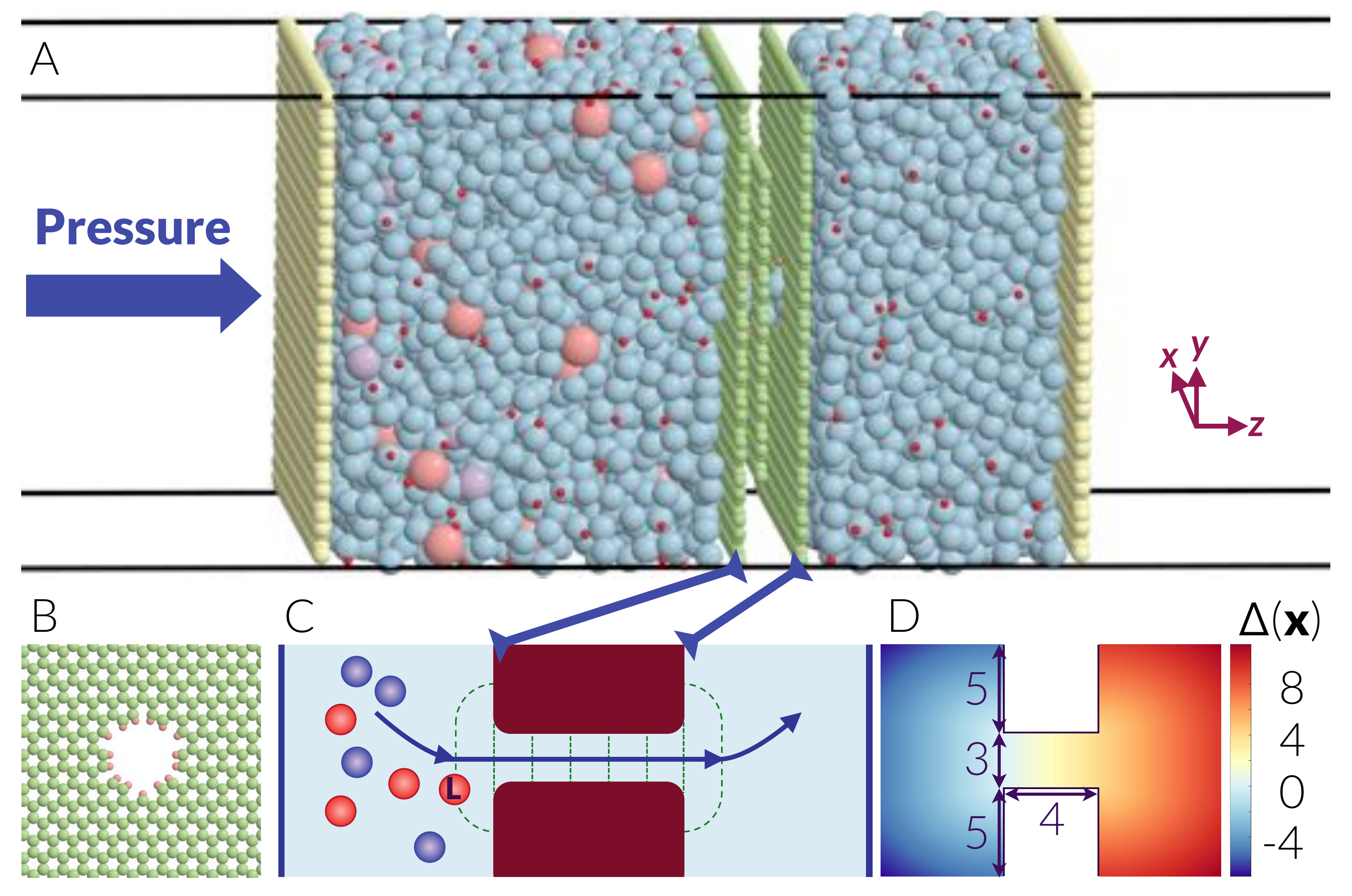}
    \caption{{\bf Overview of the Simulated System and the Employed Order Parameter:} (A-B) Schematic representation of the model filtration system with the membrane (green), two pistons (yellow), water molecules (blue {oxygens and red hydrogens}), and sodium (light purple) and chloride (peach) ions. The cross-section of the graphitic pore is depicted from above in (B) with passivating hydrogens depicted in red. Hydrostatic pressures of 195.6~bar and -0.98~bar are applied on the left and right hand side pistons, respectively. (C) Schematic depiction of $\Delta(\textbf{x})$, the directed curved distance from pore entrance, alongside with several representative milestones. $\Delta(\textbf{x})$ increases along the direction of the arrow. The order parameter is defined as the directed curved distance of the leading ion, labeled by \textbf{L}. (D) A schematic heat map of $\Delta(\textbf{x})$ for a hypothetical nanoporous membrane. The dimensions do not correspond to the membrane system considered in this work. }
    \label{fgr:Schematic}
    \vspace{-5pt}
\end{figure*}

\noindent\textbf{System Description and Preparation:} The model filtration system considered in this work comprises a three-layer graphitic membrane with a sub-nm cylindrical nanopore, two pistons, 5,720 water molecules,  95 Na$^+$ ions, and 95 Cl$^-$ ions (Figure~\ref{fgr:Schematic}A-B). The details of the simulation setup are all described in the SI. At the beginning of the simulation, the  container on the left is comprised  of $\sim$3,400 water molecules and all the ions, corresponding to an NaCl concentration of 1.5~M, while the container on the right is filled with water molecules only. The pore diameter, $0.5\pm0.2$~nm, is chosen in accordance with earlier studies of single-layer~\cite{SintJACS2008, Suk2010, CohenNanoLett2012, KonathamLangmuir2013, CohenNanoLett2016}, and multi-layer graphene~\cite{ChenACSApplMaterInt2017}, which predict considerable salt rejection for pores of comparable diameters. (See SI  for a detailed discussion of the source of uncertainty in pore diameter.) The carbon atoms within the pore interior are  all passivated with hydrogens. This choice is guided by earlier molecular simulations of desalination~\cite{CohenNanoLett2012, WangCarbon2017}, as well as the experimental realization that passivating graphene pores with hydrogen is the simplest choice for assuring their stability~\cite{LeePNAS2014}. Moreover, our choice to consider three layers of graphene with a cylindrical sub-nm pore is to assure that ion passage times are too long to be captured via conventional MD. Water molecules are represented using the \textsc{Tip3p} force-field~\cite{Price2004}. All other atoms are represented as charged  Lennard-Jones particles, with the interaction parameters and partial charges adapted from Refs.~\citenum{Muller1996, Joung2008, Beu2010} and given in Table S1. We use \textsc{Packmol}~\cite{Martinez2009} and \textsc{Lammps}~\cite{Plimpton1995} to generate and equilibrate 100 independent starting configurations using the procedure outlined in the SI.  This is to assure that our findings are not impacted by the particulars of our initial setup. Prior to being used in FFS calculations,  each configuration is  energy-minimized using the \textsc{Fire} algorithm~\cite{BitzekPRL2006} and subsequently equilibrated for a minimum of 10~ns at each state point using the non-equilibrium MD scheme described below.\\

\noindent\textbf{Molecular Dynamics Trajectories:} All MD simulations are conducted using \textsc{Lammps}~\cite{Plimpton1995}, with equations of motion integrated using the velocity Verlet algorithm~\cite{SwopeJCP1982} and a time step of 1~fs. A Nos\'{e}-Hoover thermostat~\cite{NoseMolPhys1984, HooverPhysRevA1985} with a time constant of 0.1~ps is applied to the water molecules and Na$^+$ and Cl$^-$ ions, while the carbon and hydrogen atoms within the membrane are kept fixed during the simulation. We use the \textsc{Shake} algorithm~\cite{RyckaertJCompPhys1977} to enforce the rigidity of water molecules in the \textsc{Tip3p} model.  All long-range electrostatic interactions are estimated using the particle-particle particle-mesh (\textsc{Pppm}) method~\cite{Hockney1989}, with a real-space short-range cutoff of 1.0~nm. Considering the inhomogeneity of the system along the $z$ direction and  in order to avoid well-documented artifacts arising from applying full periodic boundaries in inhomogeneous systems,  the slab \textsc{Pppm} method~\cite{BostickBiophysJ2003} is utilized in which periodic boundary conditions are only applied in the $x$ and $y$ directions.

In order to mimic the non-equilibrium nature of desalination (i.e.,~the existence of an external field, such as a hydrostatic pressure gradient), we use non-equilibrium MD~\cite{NguyenJCTC2012, CohenNanoLett2012, CohenNanoLett2016} in which an extra force $f_{h,z}$ is applied to the constituent atoms of each piston as follows. At every time step, $F_z$, the $z$ component of the total force exerted onto the $n_p=1,008$ constituent atoms of each piston is computed, and a force of $F_z/n_p+f_{h,z}$ is applied to each piston atom along the $z$ direction. The hydrostatic pressure applied to the piston will then be given by $P=n_pf_{h,z}/a_p$, with $a_p$ being the piston's surface area. This scheme is implemented in the \texttt{fix aveforce}  routine of \textsc{Lammps}, which we use in order to apply hydrostatic pressures of $195.6$ and $-0.98$~bar on the feed and filtrate pistons, respectively. In order to maintain pistons' rigidity, no force is exerted on their constituent atoms in the $x$ and $y$ directions. Also, no thermostat is applied to piston atoms and they evolve according to the microcanonical (NVE) ensemble.\\

\noindent\textbf{FFS Calculations:} Since ion transport through a nanopore is a rare event, we probe its kinetics using FFS~\cite{Allen2006}. In general, a rare event corresponds to an infrequent transition between $A$ and $B$, two (meta)stable basins within the free energy landscape of the underlying system, distinguished by an order parameter $\lambda(\textbf{x}^N)$ that quantifies the progress of transition from $A=\{\textbf{x}^N:\lambda(\textbf{x}^N)<\lambda_A\}$ to $B=\{\textbf{x}^N:\lambda(\textbf{x}^N)\ge\lambda_B\}$. FFS estimates the rate of such a transition by recursively computing the flux of trajectories that leave $A$ and cross a succession of $N$ intermediate milestones, $\lambda_A<\lambda_0<\cdots<\lambda_{N-1}<\lambda_N=\lambda_B$. Unlike most other path sampling techniques, FFS can be utilized even when the underlying dynamics is not reversible, and is therefore ideal for use with the non-equilibrium MD scheme utilized here. 
{We denote the basins of interest based on the number of each solute type in the filtrate. More precisely, $F_{p,q}$ constitutes all the configurations in which $p$ sodiums and $q$ chlorides  are present in the filtrate. The order parameter, $\lambda(\textbf{x}^N)$, is then constructed based on $\Delta(\textbf{x}_i)$'s, the directed curved distance of solute $i$ from the pore entrance. The isosurfaces of $\Delta(\textbf{x}_i)$ are schematically depicted in Fig.~\ref{fgr:Schematic}C while its precise mathematical definition is given in the SI. Intuitively, $\Delta(\textbf{x})$ is the minimum distance that it takes for an ion at $\textbf{x}$ to travel to reach the pore entrance. For a transition from $F_{p,q}$ to $F_{p,q+1}$, for instance, $\lambda(\textbf{x}^N)$ will be the $(q+1)$th largest $\Delta(\textbf{x}_i)$ for all chlorides in the system.}
We utilize jumpy FFS (jFFS)~\cite{HajiAkbariJChemPhys2018}, a generalized variant of FFS that we recently developed for order parameters that can undergo considerable changes between successive samplings, and the temporal coarse-graining approach described in Ref.~\citenum{HajiAkbariPNAS2015} with a sampling time of 0.5~ps (or 500 time steps). (See SI for a detailed technical explanation of why jFFS needs to be employed even though the utilized order parameter is mathematically continuous.) For chlorides, we compute mean passage times at five different temperatures, equally spaced between 280~K and 360~K{, while for sodium, only one calculation at 280~K is conducted}. Further details about the order parameter and the algorithm are included in the SI.

\section{Results}

\begin{figure*}
    \centering
    \includegraphics[width=.65\textwidth]{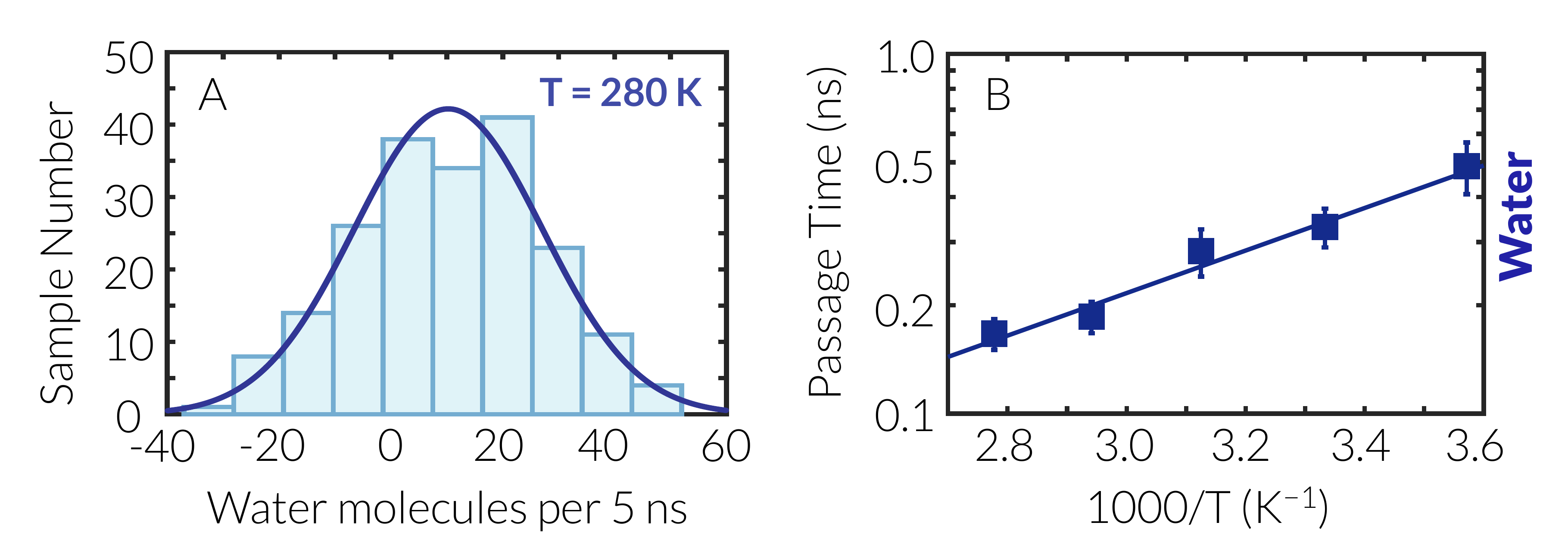}
    \caption{{\bf Kinetics of Water Transport:} (A) Statistical distribution of the number of water molecules entering the pure water container during 200 non-overlapping 5-ns windows at 280~K. The dark blue curve is a Gaussian fit to the data. (B) Arrhenius-like dependence of mean passage times for water molecules on temperature.}
    \label{fgr:Permeability}
    \vspace{-10pt}
\end{figure*}

\noindent\textbf{Water Permeability:} The passage of water molecules through semipermeable membranes is not a rare event, and its kinetics can be studied using conventional MD. We analyze the MD trajectories conducted within the starting basin as part of jFFS (See SI for details.), and compute $\Delta n_{w,p}^T=n_{w,p}(t+T)-n_{w,p}(t)$,  the change in the number of water molecules within the pure water container over a time window  $T=5$~ns. The mean passage time is then computed as $\tau_w=T/\langle\Delta n_{w,p}^T\rangle$. Note that individual $\Delta n_{w,p}^T$'s-- computed for non-overlapping windows-- exhibit considerable variability as can be seen in Fig.~\ref{fgr:Permeability}A. Obtaining an accurate estimate of $\langle\Delta n_{w,p}^T\rangle$ therefore requires analyzing trajectories initiated from a large number of independent starting configurations (100 in this work). The computed $\tau_w$'s exhibit an Arrhenius dependence on temperature with an activation energy of $\Delta{E}_{w}=11.3\pm3.4$~kJ/mol, which is considerably smaller than what has been experimentally reported for real semipermeable membranes with similar pore sizes, which span a wide range~\cite{GaryBoboJGenPhysiol1971, MehdizadehIndEngChemChem1989, KumarPNAS2007, RichardsEnvirSciTech2013}, but are generally larger than $\sim$14.2~kJ/mol~\cite{KumarPNAS2007}. This discrepancy can be attributed to the fact that in comparison to real water, transport properties depend more weakly on temperature in the \textsc{Tip3p}~\cite{Price2004} force-field. For instance, for shear viscosity, which is the most relevant transport property for pressure-driven flow within a nanopore, $\Delta{E}_{\text{visc}}^{\textsc{Tip3p}}=7.4\pm2.3$~kJ/mol (computed from the data in Ref.~\citenum{VenableJPhysChemB2010}) is almost twice smaller than the experimental value of  $\Delta{E}_{\text{visc}}^{\text{exp}}=15.7\pm0.5$~kJ/mol (computed from the data given in Ref.~\citenum{KestinJPhysChemRefData1978}).  It has indeed been argued that the activation energy for membrane permeability is bounded from below by that of transport properties of the solvent, such as shear viscosity~\cite{ChenDesalination1983}. Our computed $\Delta{E}_{w}=11.3\pm3.4$~kJ/mol is indeed {slightly} larger than{-- but statistically indistinguishable from--} $\Delta{E}_{\text{visc}}^{\textsc{Tip3p}}=7.4\pm2.3$~kJ/mol, and is consistent with other computational estimates of permeability activation energies when the \textsc{Tip3p} force-field is utilized~\cite{DahanayakaPCCP2017}.

The computed $\tau_w$'s can also be used for assessing the validity of the Hagen-Poiseuille (HP) law~\cite{SuteraAnnRevFluidMech1993}, which predicts the pressure gradient needed for maintaining a particular $\tau_w$:
\begin{eqnarray}
\Delta{P} &=& \frac{8\mu_w lM_w}{\pi\rho_w N_A\tau_w r^4}
\label{eq:HP}
\end{eqnarray}
Here, $l$ and $r$ are the length and the radius of the nanopore; $M_w, \mu_w$, and $\rho_w$ denote the molar mass, viscosity, and density of liquid water, and $N_A$ is the Avogadro constant. We utilize the viscosity and density estimates of Refs.~\citenum{VenableJPhysChemB2010} and~\citenum{MolineroJPCB2009}, respectively, and use a value of $l=0.67$~nm based on inter-layer distance of 0.355~nm in graphite. 
There is, however, some uncertainty in determining $r$  since water molecules and the nanopore interior have comparable sizes. Depending on how the accessible volume within the nanopore is defined, $r$ values as small as 0.15~nm (Fig.~S1B) and as large as 0.35~nm (Fig.~S1A) can be obtained (See SI for discussions). Considering the quartic dependence of $\Delta{P}$ on $r$, this ambiguity results in $\Delta{P}$'s that differ by a factor of 30. 
At 280~K, for instance, the estimated $\Delta{P}$ values range between 28~bar (for $r=0.35$~nm) and 830~bar (for $r=0.15$~nm). 
Despite these uncertainties, our applied pressure gradient of $\sim$196~bar falls within this range, which suggests that Hagen-Poiseuille law provides reasonable estimates of solvent flux even for a sub-nm nanopore such as the one considered here. This is also consistent with the fact that the activation energies for permeability and viscosity are statistically indistinguishable.  It is, however, necessary to note that assessing the validity of the Hagen-Poiseuille law in nanopores is a complicated proposition, due to uncertainties in determining geometric properties such as pore radius, possible breakdown of continuum assumptions at the nanoscale, such as the violation of the no-slip boundary conditions and lack of a fully developed flow within the pore. Despite these uncertainties, our calculations reveal that such deviations do not adversely impact the ability of the Hagen-Poiseuille flow to estimate the order of the magnitude of the solvent flux, as has been shown in earlier computational studies of liquid flow through nanopores~\cite{TakabaJChemPhys2007}. More precisely, we do not observe a substantial enhancement in flux as has been previously reported in experimental~\cite{HoltScience2006} and computational~\cite{GoldsmithJPhysChemLett2009} studies of water flow through carbon nanotubes. This is not surprising since molecular simulations have revealed that functionalized graphene surfaces possess substantially smaller slip lengths~\cite{WeiACSApplMaterInter2014, ChenJPhysChemC2017}. Indeed, the passivating hydrogens inside the pore can attract oxygen atoms in water, an effect that has also been shown to increase friction and decrease the flow rate in other systems~\cite{HornerScienceAdv2015}.

\begin{table*}
  \centering
  \caption{Mean first passage times for water molecules ($\tau_w$) and {chloride ions} ($\tau_s$), solute passage ratios ($S$), and total durations of MD trajectories during jFFS ($T_{\text{jFFS}}$), as a function of temperature.}
  \label{tbl:Table1}
  \begin{tabular}{cccccc}
  \hline\hline
    ~~T (K)~~  &  ~~$\tau_w$~(ns)~~ &~~$\tau_s(\mu$s)~~ & ~~$S~(\times10^{-4})$~~ &~~ $T_{\text{jFFS}}(\mu$s)~~ \\
    \hline
    280 & 0.487$\pm$0.080 & 5.63$\pm$0.34 & 0.87$\pm$0.15 & 15.00 \\
    300 & 0.331$\pm$0.041 & 2.64$\pm$0.16 & 1.25$\pm$0.17 & 9.42 \\
    320 & 0.283$\pm$0.043 & 1.78$\pm$0.11 & 1.59$\pm$0.28 & 6.96 \\
    340 & 0.186$\pm$0.018 &1.33$\pm$0.08 & 1.40$\pm$0.16 & 5.35 \\
    360 & 0.167$\pm$0.016 & 0.90$\pm$0.05 & 1.84$\pm$0.21 & 4.74 \\
    \hline
  \end{tabular}
\end{table*}

\begin{figure*}
\centering
\includegraphics[width=.79\textwidth]{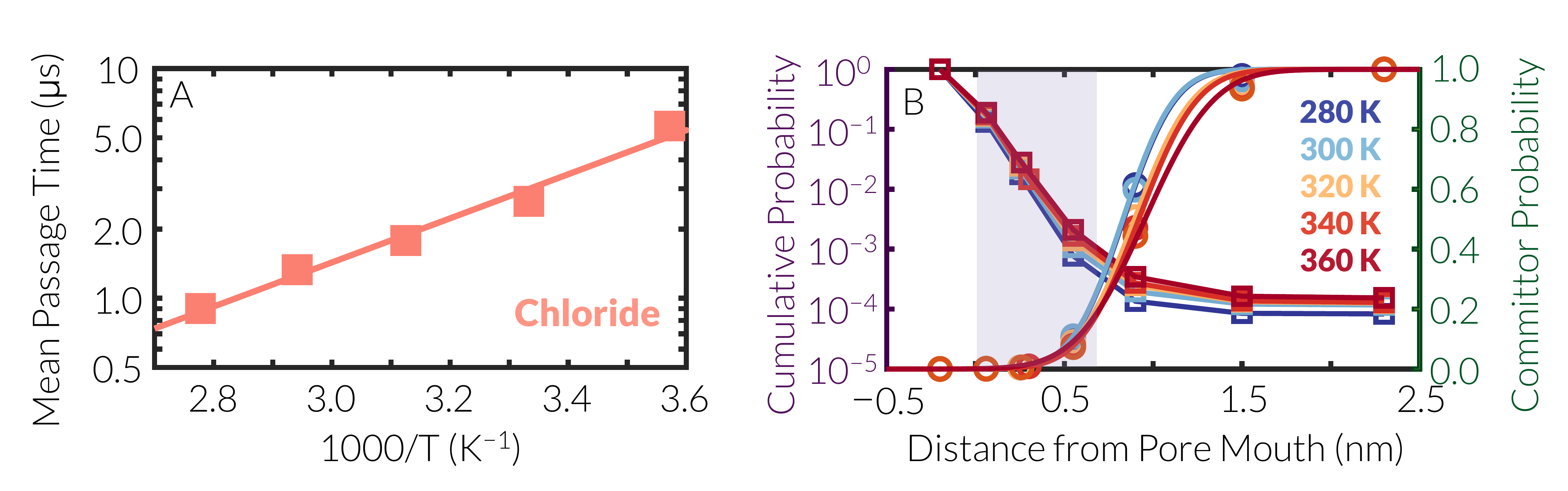}
	\vspace{-10pt}
\caption{\label{fig:selectivity}{{\bf Kinetics of Chloride  Transport:} (A) Arrhenius-like dependence of mean passage time for Cl$^-$ on temperature. (B) Cumulative transition probabilities (squares) and committor probabilities (circles) as a function of the order parameter at different temperatures. The shaded purple region corresponds to the pore interior. } }
\vspace{-10pt}
\end{figure*}

\noindent\textbf{Kinetics of Ion Transport and Selectivity:} Unlike solvent molecules, which can readily traverse the pore over sub-ns timescales, the kinetics of solute transport through nanopores is usually too slow to be accurately and efficiently captured using conventional MD. Indeed, throughout our  MD simulations within {$F_{0,0}$} (with a total duration of $\sim1~\mu$s at each temperature), we only observe one ion passage event at 360~K, and not at any other temperature. This is consistent with earlier computational studies~\cite{CohenNanoLett2012, HeiranianNatComm2015, CohenNanoLett2016, ChenACSApplMaterInt2017} of nanopores with comparable sizes, all reporting 100$\%$ salt rejection for situations under which $\tau_s$, the mean solute passage time, exceeds the duration of the conducted MD simulations. We overcome this  limitation by utilizing jFFS, which enables us to accurately and efficiently estimate arbitrarily long $\tau_s$'s. 
{In order to automatically select for the fastest passing ion, we choose our target basin as $F_{0,1}\cup F_{1,0}$ and define our order parameter as $\lambda(\textbf{x}^N) = \max_{1\le i\le n_s}\Delta(\textbf{x}_i)$, with $n_s$ the total number of ions in the system. Since $\lambda(\textbf{x}^N)$ does not distinguish between different ion types, it is suitable for exploring the $F_{0,0}\rightarrow F_{0,1}\cup F_{1,0}$ transition.}
Table~\ref{tbl:Table1} summarizes the computed solute passage times, which are on the order of microseconds. Estimating these $\mu$s-scale $\tau_s$'s with the reported level of statistical precision is still possible using conventional MD, but will require tens of long MD trajectories with a total duration of several hundred microseconds. With jFFS, this is achieved with considerably shorter trajectories, and thus at a fraction of the computational cost of conventional MD. Indeed, $T_{\text{jFFS}}$, the total duration of MD trajectories conducted within the {$F_{0,0}$} basin and between FFS milestones is never longer than five times the mean passage time.  As we will discuss later, for passage times that are {considerably longer}, using conventional MD becomes completely impractical, while our approach can directly estimate those with $T_{\text{jFFS}}$'s considerably smaller than  $\tau_s$.

An important quantity of interest in desalination is the solute passage ratio $S$, defined as $S:=\tau_w/\tau_s$. $S$ corresponds to the number of solute ions/molecules that pass the pore per every traversing water molecule {and depends on operating conditions such as temperature, pressure, and the concentration difference between the two containers}.   Table~\ref{tbl:Table1} summarizes the computed solute passage ratios which are on the order of $10^{-4}$ (or one ion per 10,000 water molecules), and correspond to a $\sim$99.99\% salt rejection. These minuscule passage ratios are the smallest nonzero values reported in the computational literature, and could not have been computed without jFFS. Yet, the ability to compute them accurately is critical to computer-aided rational design of ultra-selective membranes.

Similar to $\tau_w$, $\tau_s$ exhibits an Arrhenius dependence on temperature (Fig.~\ref{fig:selectivity}A), with an activation energy of $\Delta{E}_s=18.4 \pm 4.4$~kJ/mol. As will be discussed later, this barrier entirely corresponds to the transport of chloride ions. Note that $\Delta{E}_s$ is larger than $\Delta{E}_w$, which implies the existence of additional hindrance to the passage of solutes. We will describe the physical origins of such extra hindrance in our discussion of the molecular mechanism of solute transport.\\

\noindent\textbf{Ion Transport Mechanism:} In addition to probing the kinetics of ion transport, jFFS can provide detailed mechanistic information about the ion passage process. First, we observe that the leading ion to successfully traverse the pore is always a chloride. 
This can be attributed to {favorable interactions between the positively charged passivating hydrogens in the pore interior and the chloride ions, which decrease the free energy barrier to their crossing in comparison to the positively charged  sodiums that are repelled by the hydrogens. }
{The preferential transport of negatively charged ions through hydrogenated pores is not new and has been previously observed (e.g.,~in Ref.~\citenum{SintJACS2008}).
}

In order to identify the physical processes that culminate in the successful passage of a chloride ion,  we first focus on the cumulative transition probability as a function of the order parameter which is an indirect measure of how free energy changes with $\lambda$. (See  Eq.~S9 for the definition of the cumulative probability.) As expected, the largest drop in cumulative probability  occurs within the shaded purple region, which corresponds to the pore interior (Fig.~\ref{fig:selectivity}B). Intriguingly though, the drop in probability continues even after the ion has fully entered the pore. This can be seen more vividly in the committor probability  curves of Fig.~\ref{fig:selectivity}B, which reveal  the transition state (i.e., the collection of configurations with equal probability of proceeding towards either basin) to lie at around $\lambda=0.9$~nm, i.e.,~right after the pore exit. (See Eq.~S10 for the definition of committor probability, and Fig.~S2 for a comprehensive committor analysis, which confirms that the utilized  order parameter is a good reaction coordinate.) This implies that the free energy profile, $F(\lambda)$, is neither flat within the pore interior, nor is it  symmetric around its central dividing plane, and instead has a maximum at $\lambda=0.9$~nm. 
This is in contrast to several earlier computational studies~\cite{CorryJPhysChemB2008, SongJPhysChemB2009, RichardsSmall2012}, which report $F(\lambda)$'s that are both flat and symmetric. The discrepancy originates from, among other things, the presence of a hydrostatic pressure gradient, which breaks the reflection symmetry of the system. The potential of mean force calculations in all these earlier works, however, are conducted in the absence of such external driving forces.

\begin{figure*}[ht]
	\centering
	\includegraphics[width=.9999\textwidth]{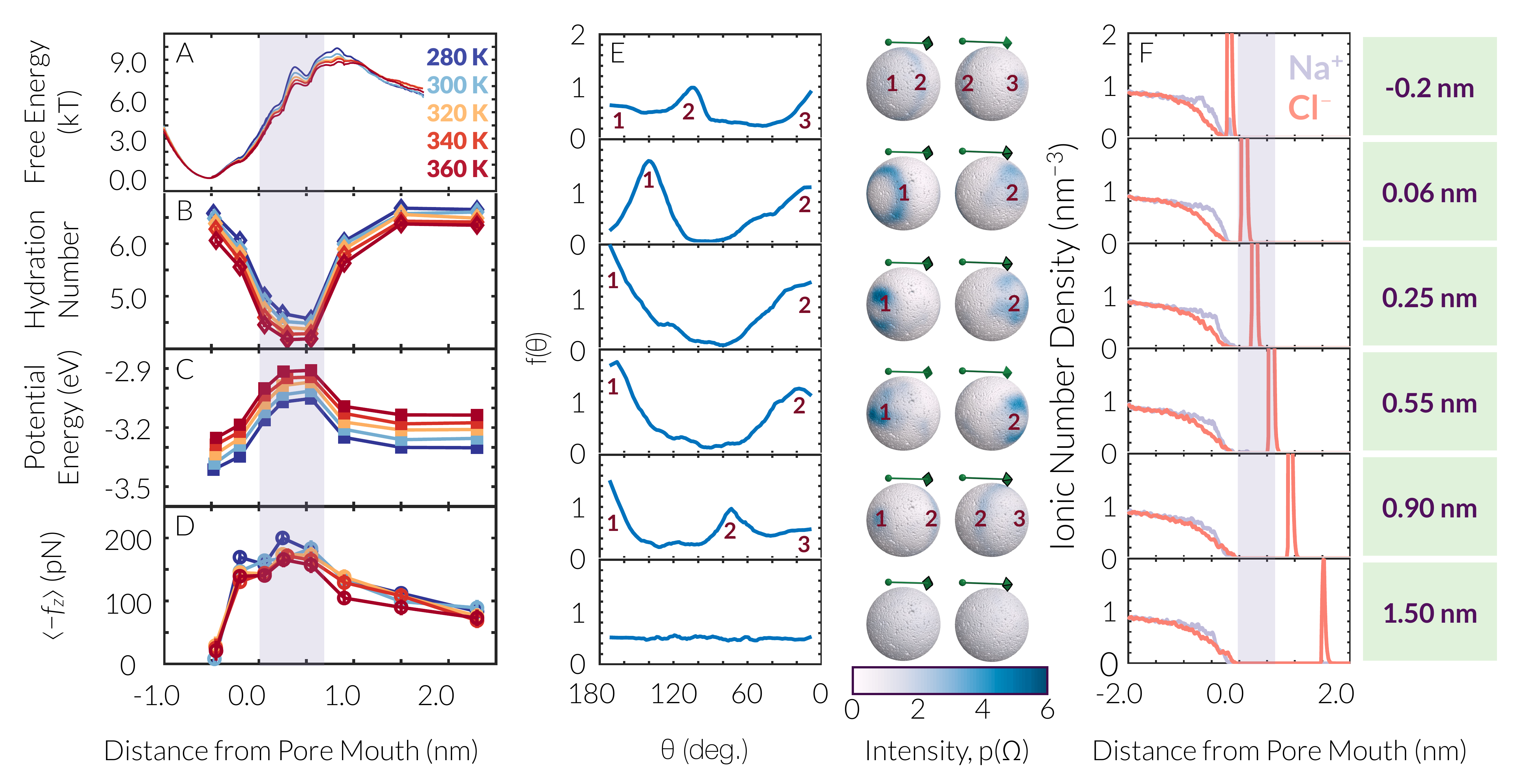}
	\caption{\label{fig:mechanism}\textbf{Mechanism of Chloride Transport through the Nanopore}~{(A) Free energy profiles computed from the FFS-MFPT method. (B)} Average hydration number, ({C}) potential energy, and ({D}) restraining force in the $z$ direction as a function of the distance from the pore mouth for the leading chloride ion at different temperatures. ({E}) The orientational distribution of water molecules within the hydration shell of the leading chloride at 280~K. Here $\theta$ is the polar angle around the $z$ axis, which is represented by the green arrows. The numbers correspond to the peaks denoted on the full spherical orientational distribution functions. {In each panel, the right and left hand side spheres are shown from the back side and front side, respectively.} ({F}) {Number density} of sodium (light purple) and chloride (dark orange) ions {as a function of curved distance from pore mouth for different values of the order parameter}. The charge {anisotropy} emerging close to the pore mouth {results in} the average restraining force along the $-z$ direction shown in ({D}). The order parameters  on the right hand side apply to both ({E}) and ({F}).}
	\vspace{-10pt}
\end{figure*}

In order to quantitatively confirm this asymmetry, we compute the free energy profiles by analyzing jFFS trajectories using the forward-flux sampling/mean first passage time (FFS-MFPT) method~\cite{ThaparJCP2015}. Even though this algorithm has been primarily developed for non-jumpy order parameters, its usage alongside jFFS is not expected to result in considerable errors since the order parameter utilized here is only slightly jumpy (Fig.~S3). As a consistency check, we compute $F(\lambda)$ from analyzing long unbiased MD trajectories in the $F_{0,0}$ basin and find no systematic difference between FFS-MFPT and MD estimates. Fig.~\ref{fig:mechanism}A depicts the free energy profiles, which have two maxima at all temperatures and are asymmetric around the central dividing plane of the nanopore. While the smaller maximum resides midway through the pore, the larger maxima coincide with the transition states determined from committor probabilities (Fig.~\ref{fig:selectivity}B). The computed barriers exhibit a weak dependence on temperature, but their average of $\Delta{E}_{s,\text{MFPT}}=21.7\pm1.1$~kJ/mol is statistically indistinguishable from the barrier estimated from the Arrhenius plot of Fig.~\ref{fig:selectivity}A, i.e.,~$\Delta{E}_s=18.4\pm4.4$~kJ/mol. This further confirms the reasonable accuracy of the FFS-MFPT approach.

In order to understand the origins of this asymmetry, we first examine the hydration properties of the leading chloride (i.e.,~the first chloride that has entered the pore). Fig.~\ref{fig:mechanism}{B} depicts its average hydration number, i.e.,~the average number of water molecules within a distance $r_c=0.375$~nm from it. Here, $r_c$ is the locus of the first valley of the chloride-oxygen radial distribution function depicted in Fig.~S4. As expected, the hydration number decreases from $\sim6$ at $\lambda=-0.2$~nm, to $\sim4.5$ at $\lambda=0.55$~nm, which coincides with a drop in cumulative probability. Due to the partial dehydration of the leading chloride during this initial stage, its potential energy increases and reaches a maximum at $\lambda=0.55$~nm (Fig.~\ref{fig:mechanism}{C}). This increase in potential energy is also accompanied by a decrease in entropy, due to structuring of the remaining water molecules within the hydration shell, as can be seen in $p(\Omega)$, the orientational distribution function for water molecules within the hydration shell (Fig.~\ref{fig:mechanism}{E}). Here, $\Omega$ is the solid angle and $\int p(\Omega)d\Omega=4\pi$. (See SI  for details.) The structuring begins even before the ion enters the pore, i.e.,~at $\lambda=-0.2$~nm where the hydration shell comprises a front peak at $\theta=0^{\circ}$ and a rear ring at the tetrahedral angle of $\theta=108^{\circ}$. At $\lambda=0.06$~nm, i.e.,~when the ion has just entered the pore, the hydration shell preserves this qualitative structure. Only the peaks become stronger and the ring is pushed back to $\theta=150^{\circ}$. As the ion proceeds through the pore, structuring becomes even more pronounced, and both the front peak and the rear ring turn into pairs of peaks at $\theta=0^\circ$ and $180^\circ$, respectively. Therefore, even though the hydration number does not change a lot within the pore interior, the hydration shell undergoes considerable reorganization. {The first-- and smaller-- maxima of the free energy profiles of Fig.~\ref{fig:mechanism}A reside within this high-energy partially-hydrated state, and are almost midway through the pore.}

As the ion leaves the pore, it gets rehydrated, its $p(\Omega)$ becomes more uniform, and its potential energy decreases. Nonetheless, its committor probability does not exceed 50\% until $\lambda=0.9$~nm, i.e.,~when the hydration number is already around six. The observed asymmetry in committor probability (Fig.~\ref{fig:selectivity}A) can therefore not be attributed to the leading ion's partial dehydration only, since all measures of hydration are symmetric around the central dividing plane as can be seen in Figs.~\ref{fig:mechanism}{B-C} and~\ref{fig:mechanism}{E}. This finding is in contrast to the traditional picture that dehydration is the primary rate limiting step in ion transport through semipermeable membranes~\cite{SahuNanoscale2017}. In order to identify alternative physical processes that cause such asymmetry, we compute $\langle\textbf{f}(\lambda)\rangle$, the average net force exerted on the leading ion as a function of $\lambda$. Unlike $\langle f_x\rangle$ and $\langle f_y\rangle$, which are statistically indistinguishable from zero (Fig.~S5), $\langle f_z\rangle$ is always negative, irrespective of temperature (Fig.~\ref{fig:mechanism}{D}). Interestingly, a net negative force of $\sim100$~pN is exerted on the leading ion long after it has left the pore. This restraining force competes with partial rehydration at the pore exit and is only overcome when the leading ion is fully rehydrated.

In order to identify the origin of this non-vanishing force, we probe the spatial distribution of sodiums and chlorides in the system while the leading chloride traverses the pore. As can be seen in Fig.~\ref{fig:mechanism}{F}, individual ions are not uniformly distributed within the feed during that process. Instead, {the number density of sodium is considerably larger than that of the chloride close to the pore mouth. Indeed, a more careful inspection of ionic distribution in the feed reveals that the first and second leading sodiums in the feed are on average closer to the pore mouth than the second and third chloride, respectively (Fig.~S6).} This charge anisotropy generates an electric field in the $z$ direction, which pulls back the leading chloride (the advancing peak in Fig.~\ref{fig:mechanism}{F}) and results in the non-vanishing $\langle f_z\rangle$ values. 
{As discussed in detail in the SI and depicted in Fig.~S7, electrostatic interactions clearly contribute to this non-vanishing $\langle f_z\rangle$, while the potential role of other factors, such as hydrodynamic interactions cannot be fully ruled out. 
}
 The anisotropy is stronger when the leading chloride is traversing the pore, and slightly weakens afterward. Nonetheless, it does not fully disappear as can be seen in Figs.~S6 and S8. This is despite the fact that prior to chloride transport (i.e.,~within the $F_{0,0}$ basin), chloride ions are more likely to be close to the pore mouth than sodiums as can be seen in Fig.~S8. However, when the leading chloride approaches and enters the pore, it induces charge anisotropy at its rear. This analysis reveals that the charge anisotropy induced by the leading ion is an important-- and overlooked-- hidden variable that controls ion transport through nanopores, and impacts both the corresponding free energy barriers and passage times.  \\

 \begin{figure*}[ht]
	\centering
	\includegraphics[width=.999\textwidth]{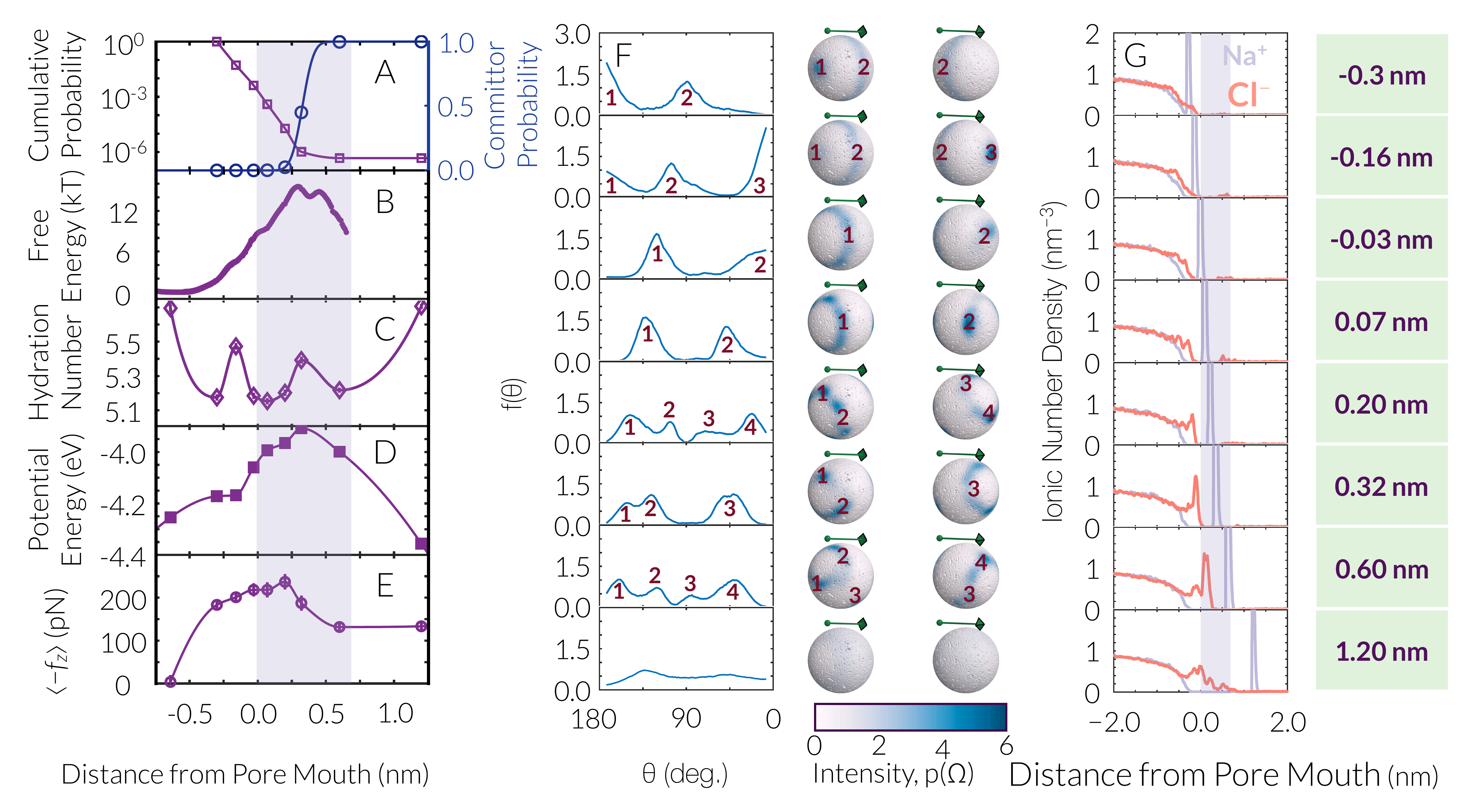}
	\caption{\label{fig:sodium}{\textbf{Kinetics and Mechanism of Sodium Transport}~(A) Cumulative transition probability and committor probability as a function of the order parameter. (B) Free energy profile computed from the FFS-MFPT method. (C) Average hydration number, (D) potential energy, and (E) restraining force in the $z$ direction as a function of the distance from the pore mouth for the leading sodium. (F) The orientational distribution of water molecules within the hydration shell of the leading sodium at 280~K. The numbers and colors are consistent with Fig.~\ref{fig:mechanism}E. (G) Number density of sodium (light purple) and chloride (dark orange) ions as a function of curved distance from pore mouth for different values of the order parameter.  The order parameters  on the right hand side apply to both (F) and (G).}}
\end{figure*}

\noindent\textbf{Sodium Transport Kinetics and Mechanism:} Since the order parameter utilized above does not distinguish between  sodiums and chlorides, it automatically selects for the ion type with the smaller passage time, namely the chloride. This means that the transition $F_{0,0}\rightarrow F_{0,1}\cup F_{1,0}$ always culminates in $F_{0,1}$. 
In order to probe the kinetics of sodium transport, we conduct FFS calculations with a more restrictive order parameter, namely $\lambda_+ :=\max_{1\le i\le{n_{\text{Na}}}}\Delta(\textbf{x}_i)$, or the maximum directed curved  distance of sodium ions from the pore mouth. Due to its high computational cost, we conduct this latter calculation at one temperature only, i.e.,~at 280~K. In order to sample the $F_{0,0}$ basin, we initiate trajectories from the endpoints of our earlier basin simulations for the $F_{0,0}\rightarrow F_{0,1}\cup F_{1,0}$ transition. Fortuitously, we find the leading chloride to  have already traversed the pore in a tiny fraction of first crossings of $\lambda_{+,0}=-0.25$~nm. This percentage, however, grows exponentially in subsequent iterations, from $\approx2\%$ at $\lambda_{+,0}=-0.25$~nm, to $\approx5.7\%$ at $\lambda_{+,1}=0.14$~nm, and $\approx34\%$ at $\lambda_{+,2}=-0.01$~nm (Fig.~S9). This implies that sodium transport is considerably faster when a chloride has already traversed the pore. The timescale for this 'faster` $F_{0,1}\rightarrow F_{1,1}$ transition is of more practical relevance since sodium transport is already expected to be considerably slower than chloride transport. We therefore stopped the calculation of the $F_{0,0}\rightarrow F_{1,0}$ transition rate, and launched a new jFFS simulation from 150 randomly selected configurations at $F_{0,1}$.  Fig.~\ref{fig:sodium}A depicts the computed cumulative and committor probabilities vs.~$\lambda_+$. The mean passage time is estimated to be $\tau_n=900\pm54~\mu\text{s}$, which is approximately 160 times longer than the passage time for chloride at the same temperature. 
 This  wide separation of timescales is consistent with the computed free energy barriers for the respective processes, i.e.,~$15.5k_BT$ for sodium (Fig.~\ref{fig:sodium}B) and $9.8k_BT$ for chloride (Fig.~\ref{fig:mechanism}A). Moreover, it is extremely costly to compute $\tau_n$ using conventional MD, while the total length of trajectories integrated during jFFS is  $T_{\text{jFFS}}\approx17~\mu$s, which is only 2\% of the $\tau_n$. This  underscores the efficiency of jFFS in estimating arbitrarily long-- and computationally inaccessible-- passage times.

In order to understand the mechanism of sodium transport, we analyze jFFS trajectories using a manner similar to chloride. There are certain aspects of the  mechanism that are shared by both ion types, such as the asymmetry of the free energy profile (Fig.~\ref{fig:sodium}B), the increase in potential energy within the pore interior (Fig.~\ref{fig:sodium}D) and the existence of a net restraining force (Fig.~\ref{fig:sodium}E). There are, however, important differences between the two processes. First of all, the transition state lies within the pore interior, i.e.,~ at $\lambda_+^*\approx0.32$~nm (Fig.~\ref{fig:sodium}A-B). This is due to the larger energetic and entropic penalty of returning to the pore for the positively charged sodium that has already reached the filtrate. Therefore, the restraining force outside the pore is not strong enough to play the same role as in chloride transport. Moreover, the hydration number only decreases slightly during the transport process (Fig.~\ref{fig:sodium}C). The fact that the leading sodium remains mostly hydrated assists it in overcoming strong electrostatic repulsions in the pore interior. Yet, even though the total number of hydrating waters does not change considerably, the hydration shell becomes highly organized as can be seen in Figs.~\ref{fig:sodium}F.  The ensuing loss of entropy partly contributes to the larger free energy barrier of Fig.~\ref{fig:sodium}B. 

Another similarity between sodium and chloride transport is the emergence of charge anisotropy in the feed (Figs.~\ref{fig:mechanism}F and ~\ref{fig:sodium}G). This effect is, however, more pronounced in the case of sodium. In particular, within a small-- but considerable-- number of configurations,  a trailing chloride  follows the leading sodium into the pore, and sometimes even all the way up to the pore exit. The presence of such a trailing chloride can facilitate the transport of the positively charged sodium. It might also be suggested that sodium transport might catalyze the passage of that trailing chloride. This is, however, very unlikely since no trailing chloride reaches the critical directed curve distance of 0.9~nm identified from Figs.~\ref{fig:selectivity}B and~\ref{fig:mechanism}A. Similar to chloride transport, the oppositely charged trailing ions are sorted in the feed as can be seen in Fig.~S10. The sorting, however, is stronger than that of  chloride depicted in Fig.~S6.  \\

\noindent\textbf{Differential Permeability and Reversal Potential:} An important consequence of the differing permeabilities of different ion types is the emergence of an electrostatic potential known as \emph{reversal potential} across the membrane~\cite{RollingsNatComm2016}. In principle, the magnitude of the reversal potential can be computed from  ionic concentrations and permeabilities using the Goldman-Hodgkin-Katz (GHK) equation~\cite{HodgkinJPhysiol1952}:
\begin{eqnarray}
\Delta\Psi &=& \frac{kT}{e}\ln\frac{p_{\text{Na}^+}[\text{Na}^+]_{\text{feed}}+p_{\text{Cl}^-}[\text{Cl}^-]_{\text{filtrate}}}{p_{\text{Cl}^-}[\text{Cl}^-]_{\text{feed}}+p_{\text{Na}^+}[\text{Na}^+]_{\text{filtrate}}}\notag
\end{eqnarray}
where $p_{\text{Na}^+}$ and $p_{\text{Cl}^-}$ are sodium and chloride permeabilities, respectively, and are inversely proportional to $\tau_n$ and $\tau_s$. Under the conditions considered here, the reversal potential is $\approx-122$~mV at 280~K, which is of the same order as experimentally reported reversal potentials for similar systems~\cite{HodgkinJPhysiol1952}. Reversal potentials are important as they are indirect measures of differing permeabilities for cations and anions. Moreover, due to the logarithmic dependence of $\Delta\Psi$ on mean passage time ratios, substantial reversal potentials only emerge when a separation of timescales exists between anion and cation transport. Our reported $\Delta\Psi$, however, is only valid for this particular feed concentration as permeabilities can in principle be affected by feed and filtrate concentrations.

\section{{Discussions and} Conclusions}

We report the first application of an advanced path sampling technique to study solute transport through nanoporous semipermeable membranes, by utilizing jumpy forward-flux sampling and non-equilibrium MD. We, in particular, probe the kinetics and microscopic mechanism of NaCl transport through a three-layer graphitic membrane with a sub-nanometer pore passivated with hydrogens. Unlike water molecules that traverse the pore over sub-nanosecond timescales, ion transport occurs over much longer timescales. The newly developed jFFS algorithm enables us to accurately and efficiently estimate mean passage times for solutes, which, in this system, are on the order of microseconds {for chlorides and milliseconds for sodiums}.  This vast separation of timescales cannot be accurately probed using conventional techniques such as molecular dynamics. Mechanistically, this separation emerges primarily due to the positive charge of the sodium ions, which are repelled by the positively charged passivating hydrogens within the pore. Therefore even though other ionic properties such as diffusivity and the size of the hydration shell can impact the extent of this separation of timescales, we expect anions to always traverse the nanopores with positively-charged interiors at higher rates than cations.

We also employ jFFS to explore the molecular mechanism of solute transport. Due to the positive charge of passivating hydrogens within the pore interior, the first ion to pass the nanopore is always a chloride. By analyzing the configurations obtained at different jFFS milestones, we observe that both the partial dehydration of the leading chloride, and charge anisotropy at the pore entrance  contribute to the free energy barrier to the transport of chloride ions.  This is in contrast to the traditional picture that considers partial dehydration as the main rate-limiting step for ion transport, and underscores the role of induced charge anisotropy as a key hidden variable.  A similar mechanism is observed for sodium transport, which involves the reorganization of its hydration shell, and the emergence of charge anisotropy in the feed. 

The induced charge anisotropy identified in this work is conceptually similar to the  concentration polarization observed in ion exchange membranes (IEMs)~\cite{KrolJMembrSci1999, TedescoJMembrSci2016}, wherein the ions that are rejected by the IEM accumulate at its surface. Unlike the transient charge anisotropies observed in this work, such concentration gradients emerge under steady-state conditions and are therefore easier to characterize experimentally. Interestingly, such polarizations are also known to slow down ion transport in pressure-driven membrane processes, and our work provides a molecular-level insight into the origin of such slowdown.

The membrane system considered in this work differs from real graphitic membranes in terms of its rigidity, and non-polarizability. These features have been previously shown to impact solvent permeability and ion selectivity~\cite{NoskovNature2004, SakiyamaNatNanotechnol2016,  MarbachNatPhys2018, GrosjeanNatComm2019}. Consequently, properties such as mean passage times, free energy barriers and restraining forces will change if more realistic representations are adopted. We, however, do not expect our key observations, namely faster transport of anions, and the emergence of charge anisotropy, to be affected by such changes. The preference for anions emanates from the existence of positively charged hydrogens inside the pore, while charge anisotropy arises due to partial ordering of trailing cations and anions. None of these effects is expected to disappear if the membrane is flexible. The effect of polarizability is also expected to be minimal, considering earlier calculations that have shown polarization effects to be insignificant in monovalent salt solutions~\cite{WernerssonJChemTheoryComput2010, KohagenJPhysChemB2015}.  On a general note, simple force-fields have a longstanding track record of accurately predicting the underlying physics of complex phenomena, such as the nucleation of colloidal~\cite{ThaparPRL2014}, polar~\cite{LiPCCP2011, HajiAkbariPNAS2015, HajiAkbariPNAS2017} or ionic~\cite{ValerianiJChemPhys2005, JiangJChemPhys2018} crystals, and physics of hydrophobicity~\cite{SumitPNAS2012, AltabetPNAS2017} and protein folding~\cite{DuanScience1998}.

As discussed earlier, the difference between the passage times of sodium and chloride is expected to result in the establishment of a reversal potential across the membrane. This reversal potential should, however,  not be confused with the charge anisotropy that is induced \emph{during} the passage process. In other words, even if a pore does not distinguish between the ions and all ion types pass at identical rates, a similar charge anisotropy is still expected to emerge when an ion is traversing the pore.

The driven ion transport process studied in this work is non-equilibrium and irreversible in nature. It might therefore be challenging to define a proper notion of a free energy for such a driven system. It is necessary to emphasize that the free energy profiles of Figs.~\ref{fig:mechanism}A and~\ref{fig:sodium}B are essentially computed based on the implicit assumption that the two containers are effectively coupled to reservoirs with constant chemical potentials. Even though this condition is not satisfied in a strict sense, the feed and filtrate concentrations do not change considerably throughout the course of the ion transport process. The obtained profiles will therefore be reasonable estimates of the free energy landscapes of the respective ion transport processes despite being computed under such a pseudo-equilibrium condition.

The theoretical models utilized for interpreting experimental findings of solute transport through membranes are generally based on the assumption that the process through which a solute enters a membrane, and its subsequent diffusive motion within the membrane are uncorrelated. Our calculations reveal that this assumption is not always valid. In other words, when an ion enters the pore, it still needs to overcome a free energy barrier, and its motion does not follow Fickian dynamics. One can therefore not use computational approaches that are based on the idea of decoupling, such as those utilized for studying other activated processes such as crystal growth in solutions~\cite{JoswiakPNAS2018, JoswiakCrystGrowthDes2018}. It must, however, be noted that such de-coupling might become possible for sufficiently long nanopores. Determining the critical pore length beyond which the docking and intra-pore diffusion processes become decoupled is an interesting question that will be addressed in future studies. One important-- and convenient-- consequence of such a decoupling will be that the passage rates will be proportional to the concentration difference between the two containers, with a proportionality constant known as permeability.  The existence of coupling in our system  makes this simple picture inaccurate possibly resulting  in a nonlinear dependence of  passage rate on concentration, a phenomenon previously observed e.g.,~for H$_2$ flow through Pd membranes~\cite{WardJMembrSci1999}. Understanding how mean passage times depend on concentration for short nanopores  will therefore be an interesting area of future exploration. 

Our calculations clearly demonstrate that there is a preferred sequence of crossing events, i.e.,~that the transport of the first chloride ion is likely to precede that of the first sodium. What is less certain, however, is whether any additional chlorides will traverse the pore in the interim, considering the wide separation of timescales between sodium and chloride transport.   Mapping out the full sequence of crossing events therefore requires determining passage times for all the relevant $F_{p,q}\rightarrow F_{p\pm1,q\pm1}$ transitions. The statistical behavior of the system can then be characterized by constructing a stochastic Markov model, and utilizing computational methods, such as kinetic Monte Carlo. Such an exploration will be useful for systematically predicting solute transport rates and reversal potentials as a function of time for  out-of-equilibrium processes such as filtration.

Our work demonstrates the power of advanced path sampling techniques such as jFFS in exploring solute transport through nanoporous membranes and provides a scalable and efficient framework for studying the structure-selectivity relationship computationally. Such investigations, however, can benefit immensely from developing more potent path sampling techniques,  more  efficient order parameters, particularly for complex membrane and pore geometries, and more realistic force-fields. 

\section*{Acknowledgements}
\noindent
We thank P. G. Debenedetti, D. Limmer and C. Ritt for useful discussions.
A.H.-A. gratefully acknowledges the support of the National Science Foundation CAREER Award (Grant No. CBET-1751971).  These calculations were performed on the Yale Center for Research Computing. This work used the Extreme Science and Engineering Discovery Environment (XSEDE), which is supported by National Science Foundation Grant No. ACI-1548562\cite{TownsCompSciEng2014}. M.E. was supported by the NSF Nanosystems Engineering Research Center for Nanotechnology-Enabled Water Treatment (EEC1449500).

\section*{Author Contributions}
\noindent
H.M., R.E., M.E., and A.H.-A. designed the research. H.M. and A.H.-A. performed the research, analyzed the data and wrote the paper.

\bibliographystyle{apsrev}

\bibliography{References}

\clearpage

\appendix 

\section{SUPPLEMENTARY INFORMATION}

\setcounter{figure}{0}
\renewcommand{\theequation}{S\arabic{equation}}
\renewcommand{\thefigure}{S\arabic{figure}}
\renewcommand{\thetable}{S\arabic{table}}

\section*{System Setup}

The filtration system of Fig.~1A is constructed as follows. The two pistons are each comprised of a single layer of $21\times24$ unit-cell graphene perpendicular to the $z$ axis, with a carbon-carbon distance of 0.1418~nm, initially placed at $z=0$ and $z=9$~nm, respectively. The pistons move along the $z$ axis during the course of the simulation due to the applied hydrostatic pressure. In order to construct the nanoporous membrane, three graphene layers with the same  number of unit cells and orientations are placed at $z=5, 5.335$ and $5.67$~nm, respectively, and each layer is properly shifted to mimic the structure of graphite. The nanopore is then created by removing the carbon atoms within each sheet that are within a circle of radius 0.45~nm from its center.  The carbon atoms at the nanopore wall are passivated by adding a sufficient number of hydrogen atoms, with a carbon-hydrogen distance of 0.08~nm as per Ref.~\citenum{Cohen2012}. After creating the pore, most of the carbon atoms within the middle graphene layer that are more than 1.16~nm farther from  pore center are removed for computational efficiency. {The removed carbon atoms have no electrostatic charges and their interactions with water molecules and sodium and chloride ions are short-range in nature. Removing them is therefore not expected to affect our findings in a systematic manner.} After constructing the pistons and the filter, the righthand side container of Fig.~1A (i.e.,~the filtrate) is filled with 2,300 water molecules that are randomly added to the region $5.95~\text{nm}\le z\le 8.75~\text{nm}$. The left-hand side container (i.e.,~the brine feed) is filled with 95 Na$^+$ ions, 95 Cl$^-$ ions and 3,400 water molecules, all randomly added to the region $0.25~\text{nm}\le z\le 4.75~\text{nm}$. All configurations are generated using \textsc{Packmol} \cite{Martinez2009} and are energy-minimized and equilibrated using \textsc{Lammps} \cite{Plimpton1995} in accordance with the procedure described in the main text. 

\section*{Force-field Parameters}

Water molecules are represented using a modified variant \cite{Price2004} of the \textsc{Tip3p} model, optimized for use with Ewald summations. For sodium and chloride ions, $sp^2$ carbons within graphene layers, and carbons and hydrogens at the pore wall, we use the parameters given in  Joung et al. \cite{Joung2008}, Beu \cite{Beu2010}, and Muller-Plathe \cite{Muller1996}, respectively. The force-fields utilized in this work are non-polarizable and cannot  account for  charge rearrangements and polarizability effects. However, the utilized partial charges ensure that  dominant Coulombic effects are captured accurately. All employed LJ parameters and partial charges  are summarized in Table~\ref{tbl:Table1}. The LJ parameters for interactions between ''unlike`` LJ sites are obtained from the Lorentz-Berthelot mixing rules.

\subsection*{Effective Pore Size}
\begin{figure*}
\centering
\includegraphics[width=.75\textwidth]{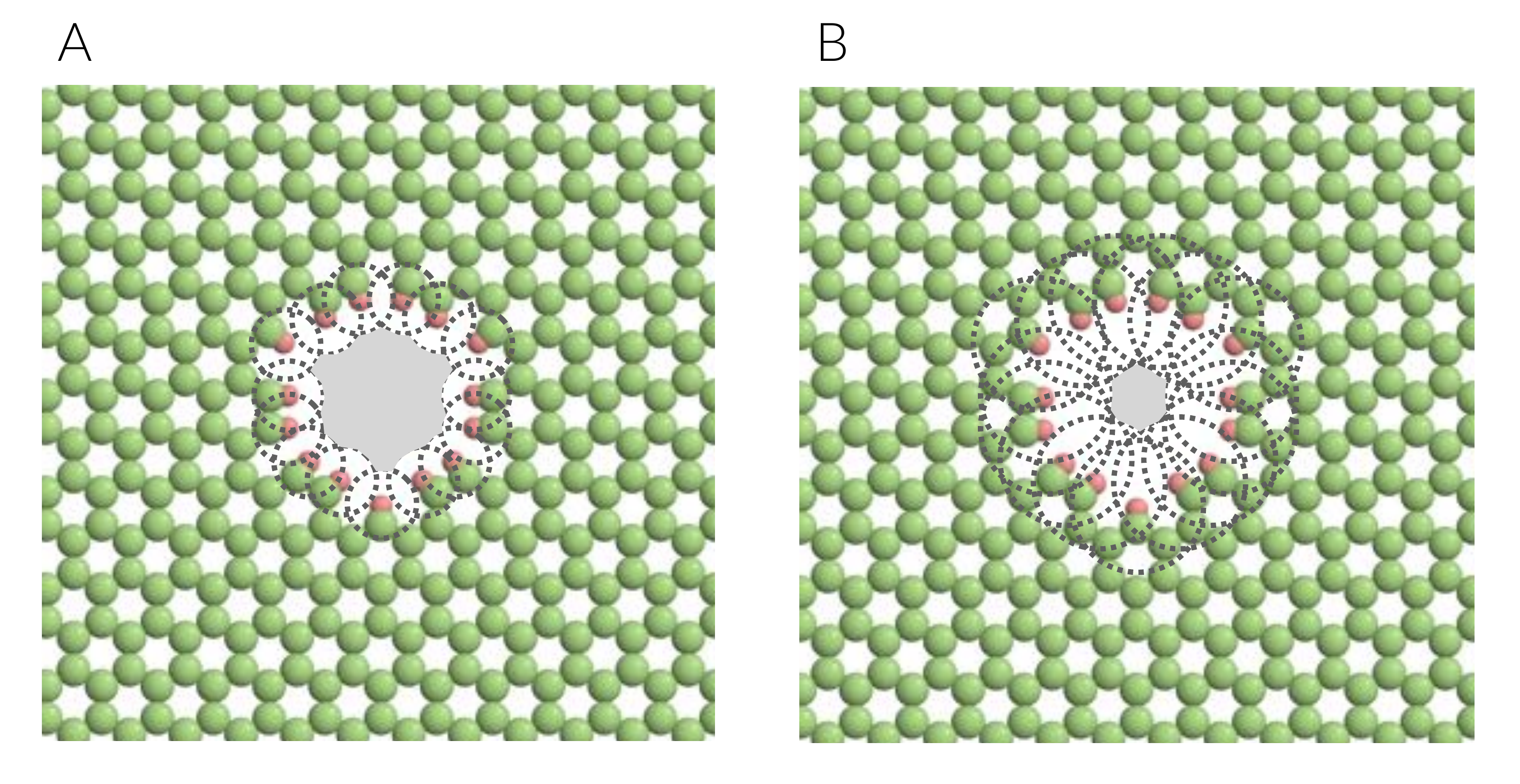}
\caption{\label{fig:pore} \textbf{Different Approaches for Defining Pore Radius:} (A) Region of the pore interior not within a distance of $\frac12\sigma_{\text{C}_{\text{CH}}-\text{O}}$ and $\frac12\sigma_{\text{H}_{\text{CH}}-\text{O}}$ from edge carbons (green) and passivating hydrogens (red). (B) Regions of the pore interior not closer than $\sigma_{\text{C}_{\text{CH}}-\text{O}}$ and $\sigma_{\text{H}_{\text{CH}}-\text{O}}$ from edge carbons and passivating hydrogens. In both cases, $a$ the area of the gray region is determined using a Monte Carlo scheme, and  the effective pore radius is determined from $r_p=\sqrt{a/\pi}$.}
\end{figure*}

Geometrically, the nanopore generated above has a diameter of $\sim0.9$~nm. The effective pore diameter, however, is much smaller and can be estimated by taking into account the van der Waals diameters of the edge carbons and passivating hydrogens. There is, however, ambiguity in defining the accessible volume within the pore. Here, we discuss two possible approaches, both using the LJ parameters for interactions between C$_{\text{CH}}$ and H$_{\text{CH}}$, and oxygens in water, as water molecules are the main entities to enter and traverse the pore. In the first approach, water molecules are treated as bulky spheres, which can only fit into the space defined in Fig.~\ref{fig:pore}A, i.e.,~the collection of points that are not within a distance $\frac12\sigma_{\text{C}_{\text{CH}}-\text{O}}$ and $\frac12\sigma_{\text{H}_{\text{CH}}-\text{O}}$ from the wall carbons and passivating hydrogens, respectively. This definition yields a pore area of $0.3997~\text{nm}^2$ and an equivalent pore radius of $r_p=$0.357~nm. The second approach defines the accessible volume as the gray area in Fig.~\ref{fig:pore}A, or the part of the nanopore that can be occupied by centers of oxygens, i.e.,~are  not within a distance $\sigma_{\text{C}_{\text{CH}}-\text{O}}$ and $\sigma_{\text{H}_{\text{CH}}-\text{O}}$ from the wall carbons and passivating hydrogens, respectively. This approach yields a pore area of $0.0715~\text{nm}^2$ and an equivalent pore radius of $r_p=$0.151~nm. As to which one of these definitions is more suitable for analyzing flow in nanopores is an open question, and has not been systematically investigated.

\section*{Forward Flux Sampling}

\subsection*{Order Parameter}

As mentioned in the main text, the order parameter $\lambda(\cdot)$ is defined based on $\Delta(\textbf{x}_i)$'s, the directed curved distance of solute $i$ from the pore mouth. In order to compute $\Delta(\textbf{x}_i)$ for a pore with a fixed-- but arbitrarily shaped-- cross section, we first construct a density map $\rho(\textbf{r})$ using the method proposed by Willard and Chandler for determining instantaneous liquid-gas interfaces \cite{WillardJPhysChemB2010}:
\begin{eqnarray}
\rho(\textbf{r}) &=& \frac1{\left(2\pi\right)^{3/2}}\sum_{i=1}^{n_m} \frac{m_i}{\sigma_i^3}\exp\left[-\frac{|\textbf{r}-\textbf{y}_i|^2}{2\sigma_i^2}\right] 
\end{eqnarray}
Here, $m_i$, $\textbf{y}_i$ and $\sigma_i$ are the mass,  position and the width of Gaussian noise of the $i$th atom in the membrane, respectively, and $n_m$ is the total number of membrane atoms. We utilize a constant value of $\sigma_i = 0.1$~nm, while $m_i$ is chosen to be the atomic mass of $i$. We compute $\rho(\textbf{r})$ over a $g_x\times g_y\times g_z$ cuboidal grid ($g_x=260,~g_y=257,~g_z=1450$), and define a two-dimensional density projection $\tilde{\rho}_{pq}$  as:
\begin{eqnarray}
\tilde{\rho}_{pq} &=& \left\{
\begin{array}{ll}
\max_{1\le r\le g_z}\rho_{pqr} & \max_{1\le r\le g_z}\rho_{pqr}\ge\rho_b\\
0 & \text{otherwise}
\end{array}
\right.\notag\\
\rho_b &=& \frac{\max_{p,q,r}\rho_{pqr}+\min_{p,q,r}\rho_{pqr}}{2}
\end{eqnarray}
Here, $\rho_b$ is the threshold for distinguishing the grid points that are part of the membrane from those within the liquid. 
The next step is to define a proximity projection map, $c_{pq}$:
\begin{eqnarray}
c_{pq} &:=& \min_{r,s,\tilde{\rho}_{rs}=0} d[(x_p,y_q),(x_r,y_s)]
\label{eq:c-pq}\\
(\xi_{pq},\eta_{pq}) &:=& \text{argmin}_{r,s,\tilde{\rho}_{rs}=0} d[(x_p,y_q),(x_r,y_s)]
\label{eq:c-pq-argmin}
\end{eqnarray}
which measures the closest distance between $(x_p,y_q)$ and a grid point with a vanishing $\tilde{\rho}$. Here $d[(x_p,y_q),(x_r,y_s)]$ is the Euclidean distance between $(x_p,y_q)$ and $(x_r,y_s)$, and $\xi_{pq}$ and $\eta_{pq}$ are the values of $r$ and $s$ that minimize Eq.~(\ref{eq:c-pq}).  Finally, we use $\rho(\textbf{r})$ to determine $z_{m}$ and $z_M$, the minimum and maximum $z$'s for grid points that are part of the membrane, i.e.,~those with $\rho(\textbf{r})\ge\rho_b$. Since the membrane geometry does not change over time, and is thus identical for all trajectories, $\tilde{\rho}, c, \xi, \eta, z_m$ and $z_M$ are all computed only once, at the beginning of each simulation, and are stored for future use.

In order to compute $\Delta(\textbf{x}_i)$ for solute $i$, we first identify $p_i$ and $q_i$, the indices for the $xy$ projection of $\textbf{x}_i$. $\Delta(\textbf{x}_i)$ will then be given by:
\begin{widetext}
\begin{eqnarray}
\Delta(\textbf{x}_i) &=& \left\{
\begin{array}{ll}
-\sqrt{d^2[(x_{p_i},y_{q_j}),(x_{\xi_{ij}},y_{\eta_{ij}})]+(z_i-z_m)^2} & z_i<z_m\\
z_i-z_m & z_m\le z_i\le z_M\\
z_M-z_m+\sqrt{d^2[(x_{p_i},y_{q_j}),(x_{\xi_{ij}},y_{\eta_{ij}})]+(z_i-z_M)^2} & z_i>z_M
\end{array}
\right.\notag\\&&
\end{eqnarray}
\end{widetext}

\subsection*{Details of jFFS}
In principle, the order parameter introduced above is a continuous function of the solute coordinates, and is not jumpy. In accordance with the coarse-graining scheme of Ref.~\citenum{HajiAkbariPNAS2015}, however, we compute the order parameter every 0.5~ps (i.e.,~every 500 MD steps). Therefore, even though $\lambda(\textbf{x})$ might not change a lot over a single MD time step, it might undergo considerable changes over one sampling window, namely 500 MD steps. {This can, for instance, be seen in Fig.~\ref{fig:op-jump}, which shows the jump probability density over 0.5-ps windows for chloride transport at 280 K.} Furthermore, it is customary to discretize a continuous order parameter for bookkeeping purposes, and if the bin size is small (e.g.,~the $\Delta\lambda=0.01$~nm utilized in this work), the order parameter can jump over several bins within a single sampling window.  We therefore use jFFS to systematically take into account such fluctuations. The technical details of jFFS are outlined in our earlier publication \cite{HajiAkbariJChemPhys2018}, with the loci of individual milestones given in Table~\ref{tab:milestones}. The algorithm utilized here is based on the procedure described in Section III B 1 of Ref.~\citenum{HajiAkbariJChemPhys2018} and works as follows. The first stage of jFFS involves conducting several MD trajectories (in our case, 100) within the $A$ basin, and enumerating the number of times that they cross $\lambda_0$ after leaving $A$. We then evaluate $\lambda_{0,\max}$, or the largest value of the order parameter for the configurations arising from such crossings, and choose $\lambda_1$ to be larger than $\lambda_{0,\max}$.  The total number of crossings divided by the total length of trajectories yields $\Psi_{A\rightarrow0}$, the flux of trajectories that end up in the interval $[\lambda_0,\lambda_1)$ upon crossing $\lambda_0$ after leaving $A$. 
{For the $F_{0,0}\rightarrow F_{0,1}\cup F_{1,0}$ transition, 100 independent trajectories are conducted in the $A=F_{0,0}$ basin, which}
 amount to a minimum of one microsecond, and result in a minimum of $\approx2,300$ crossings. 
{For sodium transport, 150 independent trajectories amounting to a total of one microsecond are conducted in the $A=F_{0,1}$ basin, and result in a total of 2300 crossings at 280~K.
}
 The next stage is comprised of $N=6$ iterations {for the $F_{0,0}\rightarrow F_{0,1}\cup F_{1,0}$ transition and $N=7$ for the $F_{0,1}\rightarrow F_{1,1}$ transition} aimed at computing the minuscule probability of reaching $\lambda_N$ from $\lambda_0$. During the $k$th iteration, a large number of MD trajectories are initiated from the configurations that have been obtained upon crossing $\lambda_{k-1}$. At the onset of each trajectory, the momenta of mobile atoms and/or molecules are randomized in accordance with the Boltzmann distribution. Each trajectory is then terminated when it crosses $\lambda_k$ or returns to $A$. Similar to basin simulations, $\lambda_{k,\max}$, the maximum value of the order parameter for configurations arising from such crossings is evaluated, and $\lambda_{k+1}$ is chosen to be larger than $\lambda_{k,\max}$. The fraction of trajectories that result in a successful crossing is the transition probability between $\lambda_{k-1}$ and $\lambda_k$, and is denoted by $\langle U_{k-1,k}\rangle$. The cumulative flux of trajectories leaving $A$ and reaching $B$ is given by:
\begin{eqnarray}
\Phi_{A\rightarrow B} &=& \Psi_{A\rightarrow0}\prod_{k=1}^N\langle U_{k-1,k}\rangle
\end{eqnarray}
The mean passage time is given by:
\begin{eqnarray}
\tau_s &=& \frac{1}{\Phi_{A\rightarrow B}}
\end{eqnarray}
In addition to total fluxes and mean passage times, the individual transition probabilities can be utilized to estimate the partial cumulative probability $P(\lambda_i|\lambda_0)$ and the committor probability $p_C(\lambda_i)$:
\begin{eqnarray}
P(\lambda_i|\lambda_0) &=& \prod_{k=0}^{i-1}\langle U_{k,k+1}\rangle\\
p_C(\lambda_i) &=& \prod_{k=i+1}^N \langle U_{k-1,k}\rangle
\end{eqnarray}
which describe the probability of reaching $\lambda_i$ from $\lambda_0$, and  reaching $B$ from $\lambda_i$ (before returning to $A$), respectively. 
In this work, we terminate each iteration after a minimum of 2,000 crossings, with 2,500--3,000 crossings required at the first two milestones. Since the target milestone for each iteration is chosen after finishing the previous iteration, slightly different values of $\lambda_k$ are utilized at different temperatures.

\section*{Orientational Distribution Function, $p(\Omega)$}
The orientational distribution functions of Fig.~4D are computed as follows. First, a uniform grid of $m=10,000$ points is generated on the surface of the unit sphere. This is achieved by generating $3m$ standard normal random numbers $\{u_{i,j}\}_{1\le i\le m}^{1\le j\le 3}$ and letting $\textbf{s}_i=\textbf{u}_i/|\textbf{u}_i|$ with $\textbf{u}_i\equiv(u_{i,1},u_{i,2},u_{i,3})$. Then, $\{\textbf{r}_i\}_{i=1}^{n_w}$, the vectors connecting the leading chloride to the oxygens of all its hydrating water molecules are computed for all configurations collected during a jFFS iteration. Each vector is represented as a Gaussian cloud, and its contribution to each grid point is enumerated accordingly. 
\begin{eqnarray}
f_i &=& \sum_{j=1}^{n_w} \exp\left[-\frac{\left|\textbf{s}_i-\frac{\textbf{x}_j}{|\textbf{x}_j|}\right|^2}{\sigma^2}\right]
\end{eqnarray}
In this work, we use a value of  $\sigma=0.13$ for the width of the Gaussian cloud. Since these $m$ points are uniformly distributed on the surface of the unit sphere, the orientational probability density for point $i$ will be given by:
\begin{eqnarray}
p_i &:=& \frac{f_i}{\sum_{j=1}^mf_j}
\end{eqnarray}
After computing $p(\Omega)\equiv p(\theta,\phi)$, $f(\theta)$ is computed as:
\begin{eqnarray}
f(\theta) = \frac1{4\pi}\int_0^{2\pi} p(\theta,\phi)d\phi
\end{eqnarray}
In other words, $\int_0^\pi f(\theta)\sin\theta d\theta=1$.

{
\section*{Ionic Number Densities vs. $\Delta$}
In order to compute the ionic number densities depicted in Figs.~4F, 5G and~\ref{fig:ion-dist-basin}, we first evaluate $v(\Delta)$, the volume of all grid cells that have the same $\Delta$ value. We then evaluate $N_i(\Delta)$ the average number of ions of type $i$ present in the grid points sharing the same $\Delta$. The number density for ion $i$ is given by $\rho_i(\Delta):=N_i(\Delta)/v(\Delta)$. 

\vspace{20pt}
\section*{Elucidating the Role of Electrostatics in the Negative Restraining Forces of Fig. 4D}

Fig.~\ref{fig:elec-cont-Cl}A depicts the electrostatic and non-electrostatic contributions to $\langle f_z\rangle$ for chloride transport at 280~K. The electrostatic force is always negative except at $\lambda=0.9$~nm. This is, however, misleading since the electrostatic force exerted on an ion is strongly dominated by the water molecules within its first hydration shell. This effect is averaged out when water molecules are uniformly distributed within the hydration shell without any preferential orientation. When the leading chloride traverses the pore, however, the hydration shell is structured and orientationally anisotropic. The contribution of the first hydration shell to the net electrostatic force can be readily calculated from $p(\Omega)$ and the average hydration number by assuming that water molecules are perfect dipoles and are all located at the first peak of the Cl-O radial distribution function: 
\begin{eqnarray}
\langle\textbf{F}_{\text{hydration}}\rangle &=& \frac{ze^2\Delta r\langle N_{\text{hyd}}\rangle}{4\pi^2\epsilon_0 r_c^3}\int_{\mathcal{S}} \textbf{e}_r(\Omega) f(\Omega)d\Omega
\end{eqnarray} 
Here, $e=1.60217662\times10^{-19}$~C is the electron charge, $z=0.417$ is the partial charge of hydrogen atoms in the \textsc{TIP3P} model, $\Delta{r}=0.0586$~nm is the  OH distance projected onto the central dividing plane of the water molecule, $r_c=0.375$~nm is the first peak of the Cl-O radial distribution function, and $\textbf{e}_r=(\sin\theta\cos\phi, \sin\theta\sin\phi,\cos\theta)$ is the unit radial vector in the spherical coordinates. By subtracting $\langle F_{\text{hydration},z}\rangle$ from the electrostatic force, we obtain a contribution that is always negative along the pore (Fig.~\ref{fig:elec-cont-Cl}B), but is always smaller in magnitude than the total restraining force. This clearly demonstrates that electrostatic charges partly contribute to the restraining forces depicted in Fig.~4D, but the role of non-electrostatic effects cannot be fully ruled out.
}

\begin{table*}
\centering
  \caption{LJ parameters and partial charges employed in this work}
  \label{tbl:Table1}
  \begin{tabular}{llllllll}
    \hline
    Element & C(sp2) & C$_{\text{CH}}$ & H$_{\text{CH}}$ & H$_w$ & O$_w$ & Na$^+$ & Cl$^-$  \\
    \hline
    $\varepsilon \, (kcal/mol)$ & 0.0859 & 0.046 & 0.0301 & 0 & 0.102 & 0.1684 & 0.0117 \\
    $\sigma \, (\AA)$ & 3.3997 & 2.985 & 2.42 & 0 & 3.188 & 2.2589 & 5.1645 \\
    q (e) & 0 & $-0.115$ & $+0.115$ & $+0.417$ & $-0.834$ & $+1$ & $-1$ \\
    \hline
  \end{tabular}
\end{table*}

\begin{table*}
\centering
  \caption{jFFS Milestones. All $\lambda_i$'s are in nanometers.}
  \label{tab:milestones}
  \begin{tabular}{lcccccccccc}
    \hline
    Transition & $T(\text{K})$ & $\lambda_A$ & $\lambda_0$ & $\lambda_1$ & $\lambda_2$ & $\lambda_3$ & $\lambda_4$ & $\lambda_5$ & $\lambda_6$ & $\lambda_7$\\
    \hline
    $F_{0,0}\rightarrow F_{0,1}\cup F_{1,0}$& 280 & $-0.50$ & $-0.20$ & 0.06 & 0.25 & 0.55 & 0.90 & 1.50 & 2.30 & --\\
    $F_{0,0}\rightarrow F_{0,1}\cup F_{1,0}$&300 & $-0.50$ & $-0.20$ & 0.06 & 0.25 & 0.55 & 0.90 & 1.50 & 2.30 & --\\
    $F_{0,0}\rightarrow F_{0,1}\cup F_{1,0}$&320 & $-0.50$ & $-0.20$ & 0.06 & 0.25 & 0.55 & 0.90 & 1.50 & 2.30 & --\\
    $F_{0,0}\rightarrow F_{0,1}\cup F_{1,0}$&340 & $-0.50$ & $-0.20$ & 0.06 & 0.30 & 0.55 & 0.90 & 1.50 & 2.30 & --\\
    $F_{0,0}\rightarrow F_{0,1}\cup F_{1,0}$&360 & $-0.50$ & $-0.20$ & 0.06 & 0.26 & 0.55 & 0.90 & 1.50 & 2.30 & --\\
    $F_{1,0}\rightarrow F_{1,1}$ & 280 & $-0.55$ & $-0.30$ & $-0.16$ & $-0.03$ & $0.07$ & $0.20$ & $0.32$ & $0.60$ & $1.20$ \\
    \hline
  \end{tabular}
\end{table*}

\begin{figure*}[h]
\centering
\includegraphics[width=.4\textwidth]{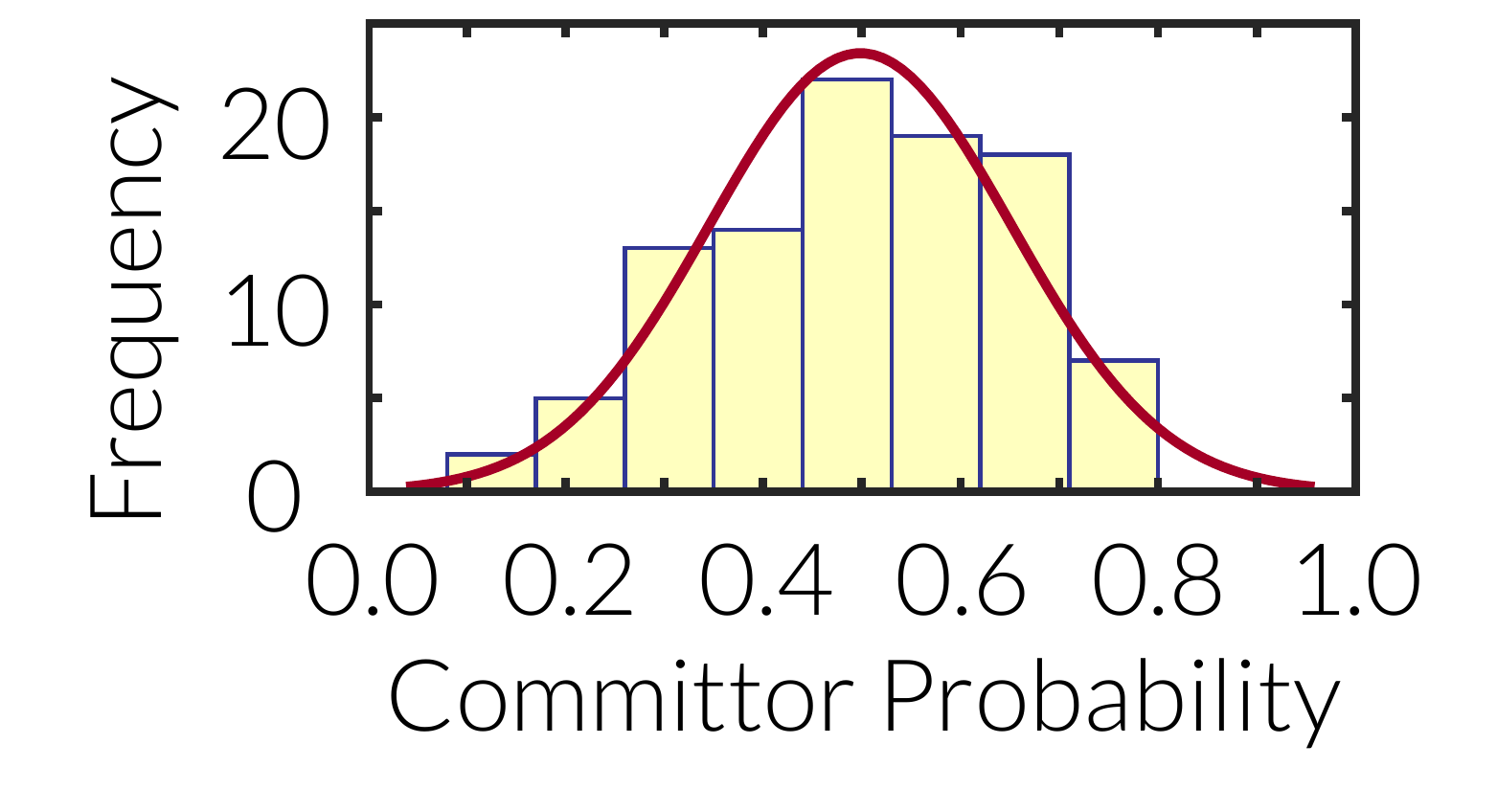}
\caption{{Committor analysis for chloride transport at 320~K. 100 configurations are randomly selected at or around $\lambda=$0.9~nm and 20 MD trajectories with randomly chosen momenta are initiated from each configuration. The per-configuration committor probability has a Gaussian distribution with mean 0.4985 and standard deviation 0.1533.}}
\end{figure*}

\begin{figure*}[h]
\centering
\includegraphics[width=.72\textwidth]{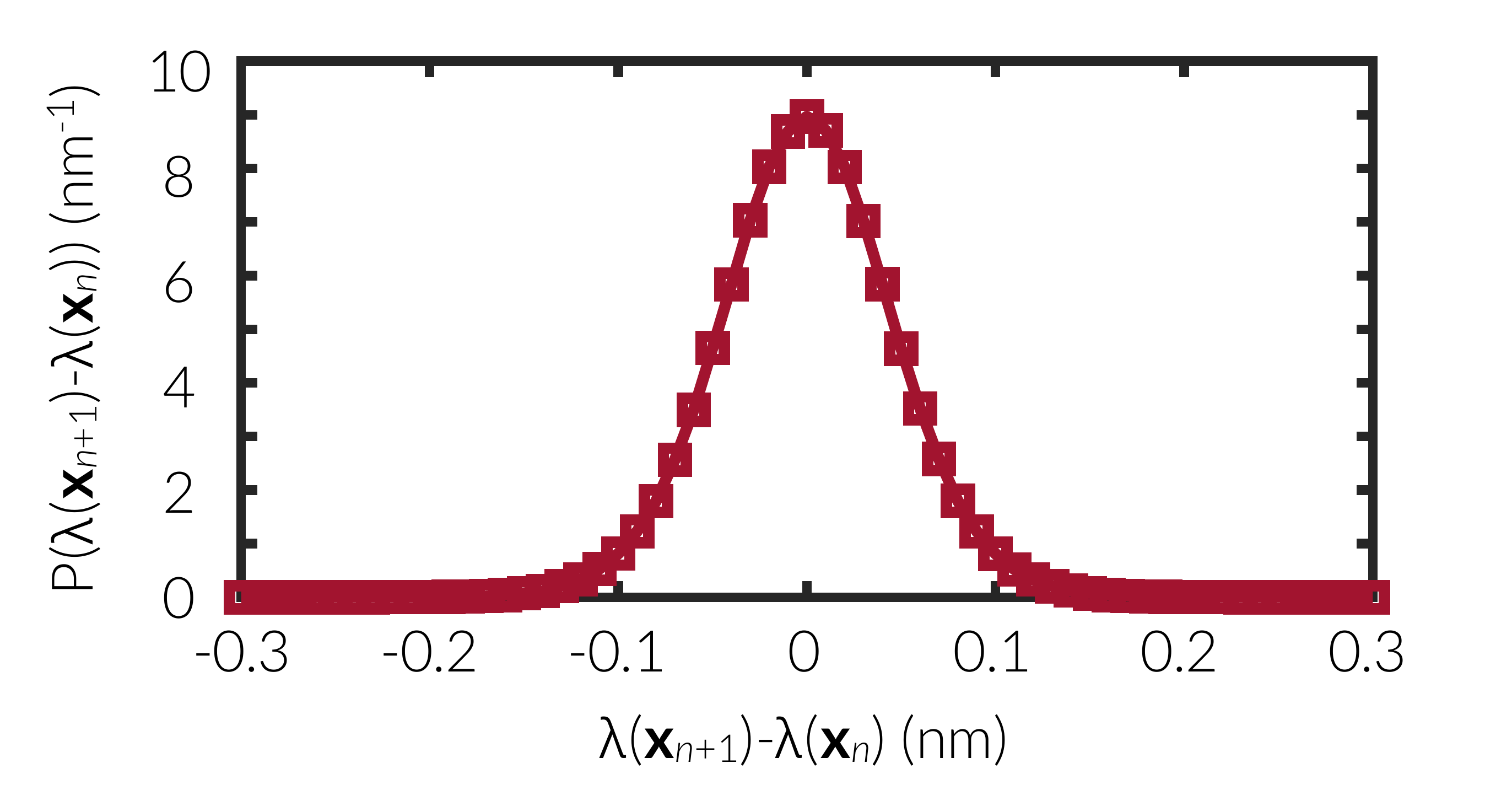}
\caption{\label{fig:op-jump}{Probability density function for jumps in $\lambda(\textbf{x})$ at 280~K over 0.5~ps time intervals, computed from trajectories employed for sampling the $F_{0,0}$ basin.}}
\end{figure*}

\begin{figure*}[h]
\centering
\includegraphics[width=.74\textwidth]{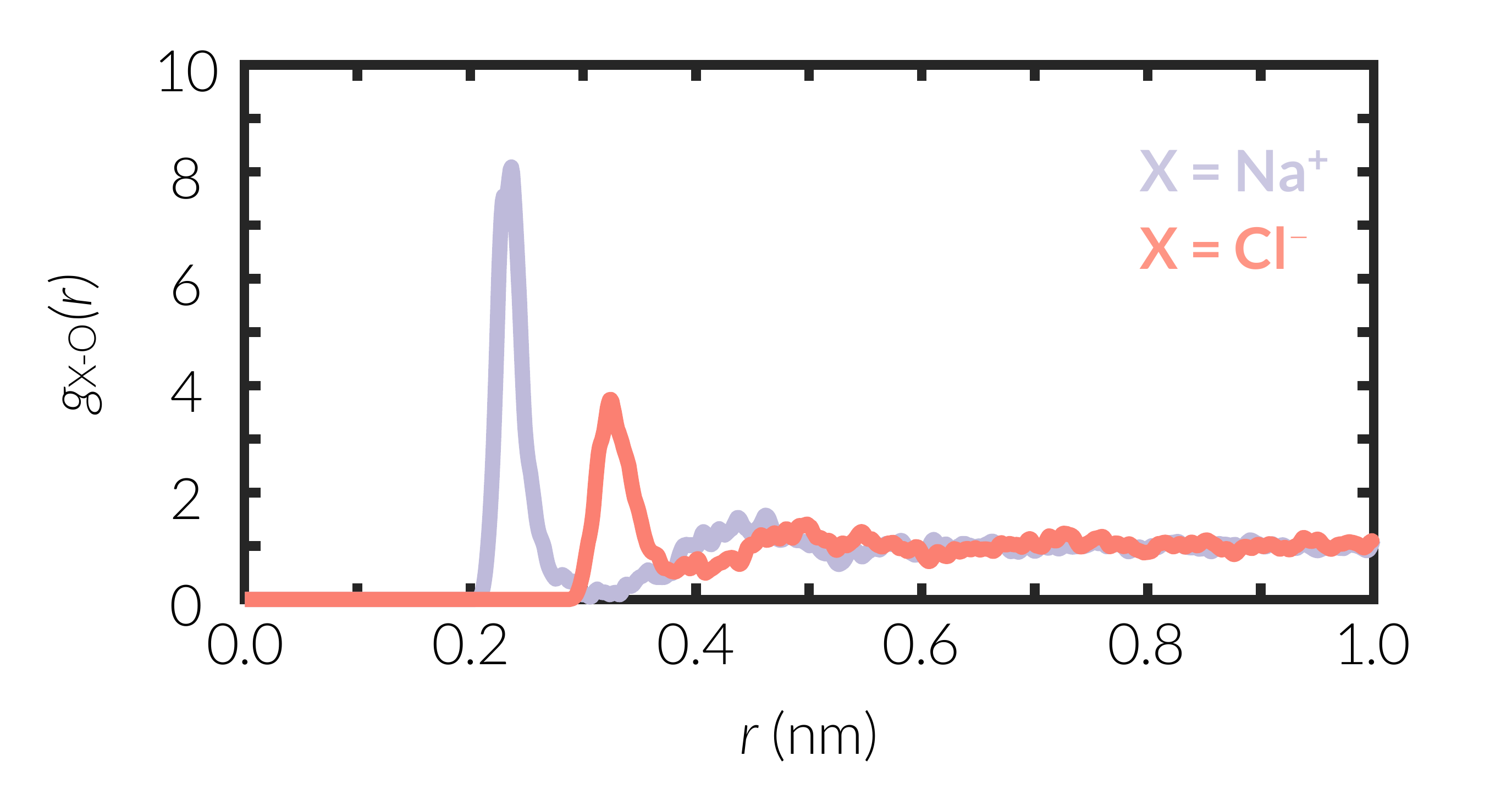}
\caption{\label{fig:rdf} Chloride-oxygen and sodium-oxygen radial distribution functions computed from a 10-ns long NPT simulation at 300~K and 1~bar.}
\end{figure*}

\begin{figure*}[h]
\centering
\includegraphics[width=.46\textwidth]{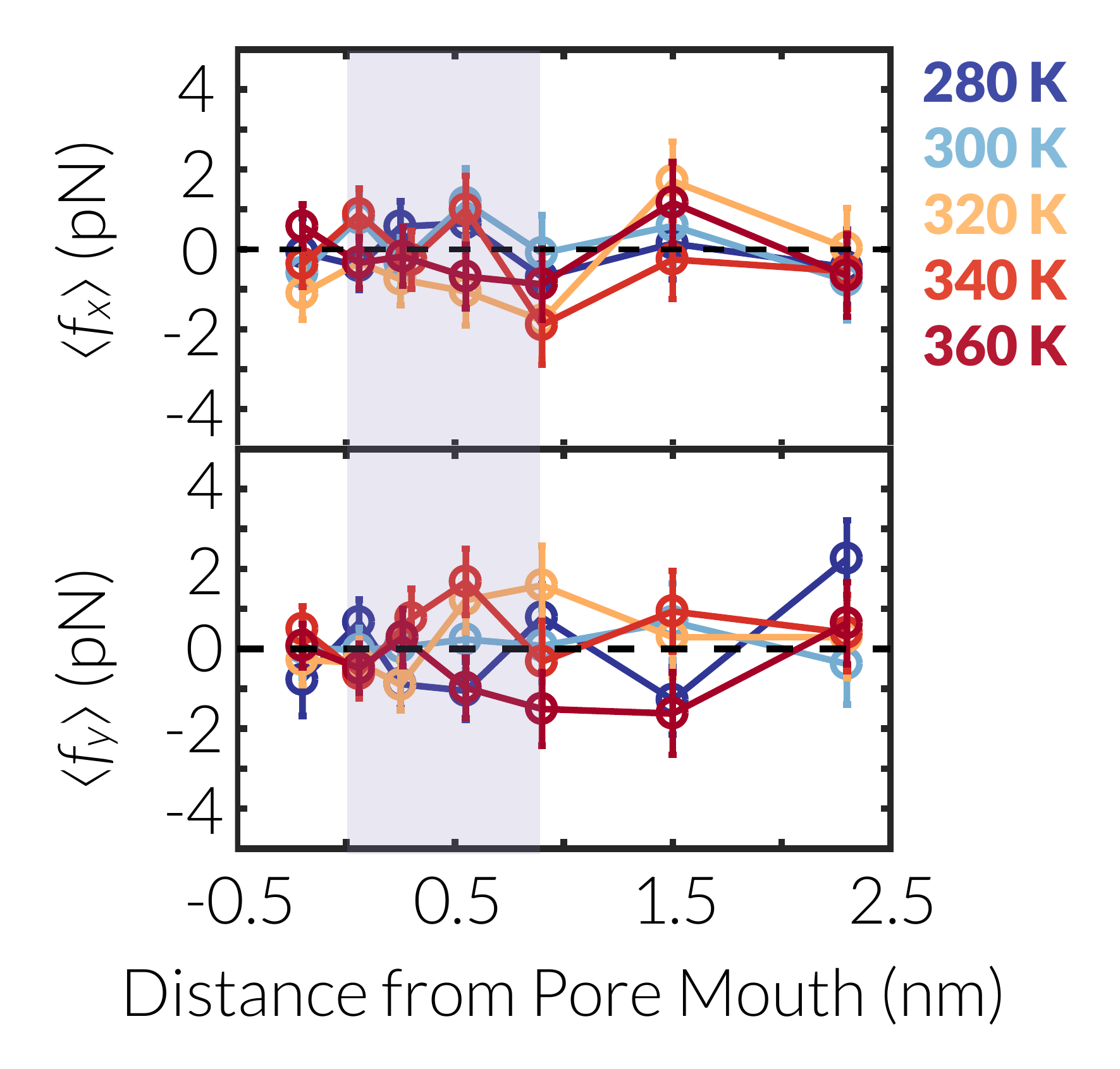}
\caption{\label{fig:force} Average force exerted on the leading chloride in the $x$ and $y$ directions as a function of distance of the leading ion from pore mouth. Note that $\langle f_x\rangle$ and $\langle f_y\rangle$ are statistically indistinguishable from zero, unlike $\langle f_z\rangle$ that is considerably larger than zero (Fig.~4D). The shaded purple region corresponds to the pore interior.
}
\end{figure*}

\begin{figure*}[h]
\centering
\includegraphics[width=.89\textwidth]{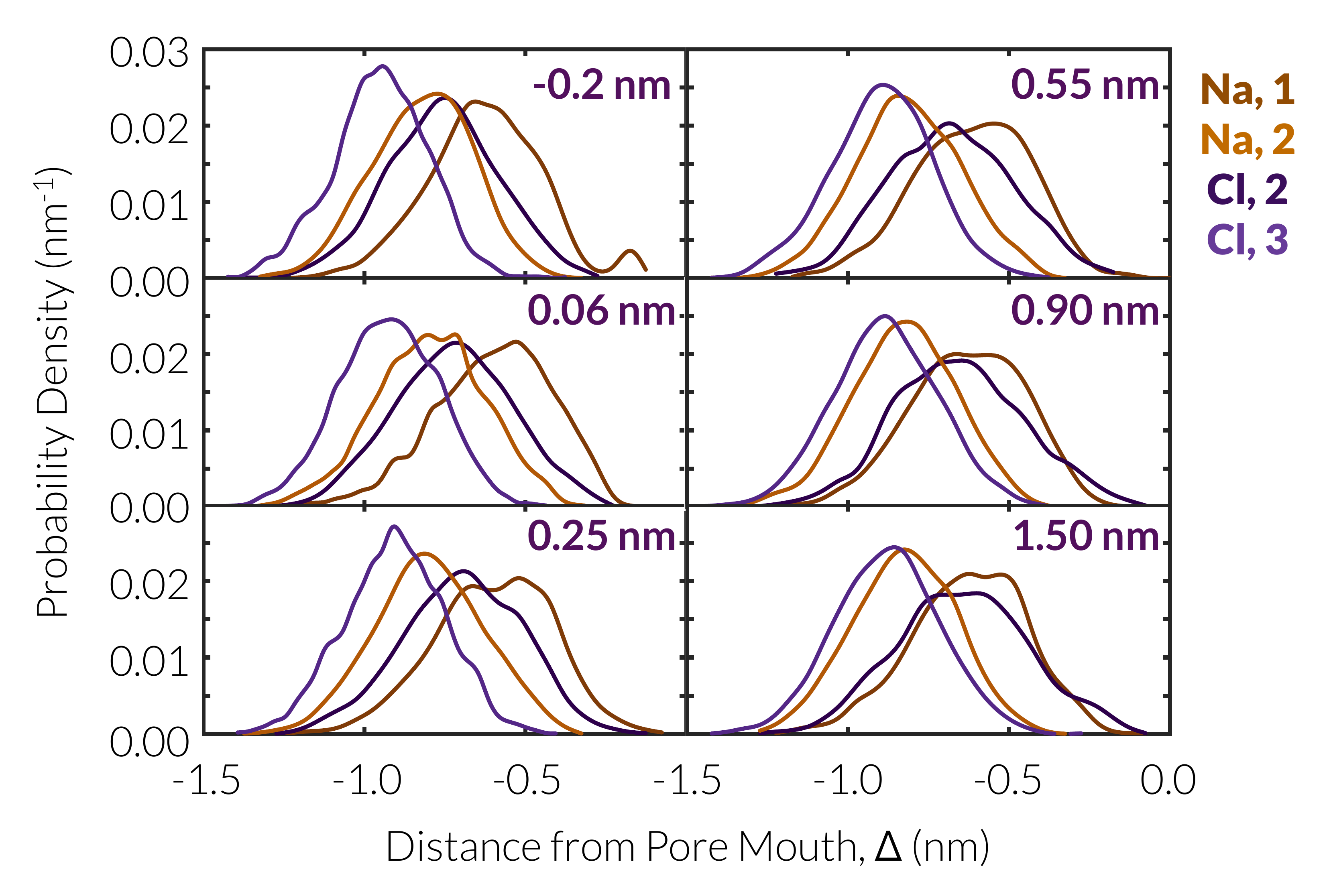}
\caption{{Spatial distribution of non-leading ions during the process of chloride transport at 280~K. The separation between the first sodium and the second chloride persists throughout the transport process and only fizzles out at 1.5~nm, i.e.,~after the transition state.} }
\end{figure*}

\begin{figure*}[h]
\centering
\includegraphics[width=.78\textwidth]{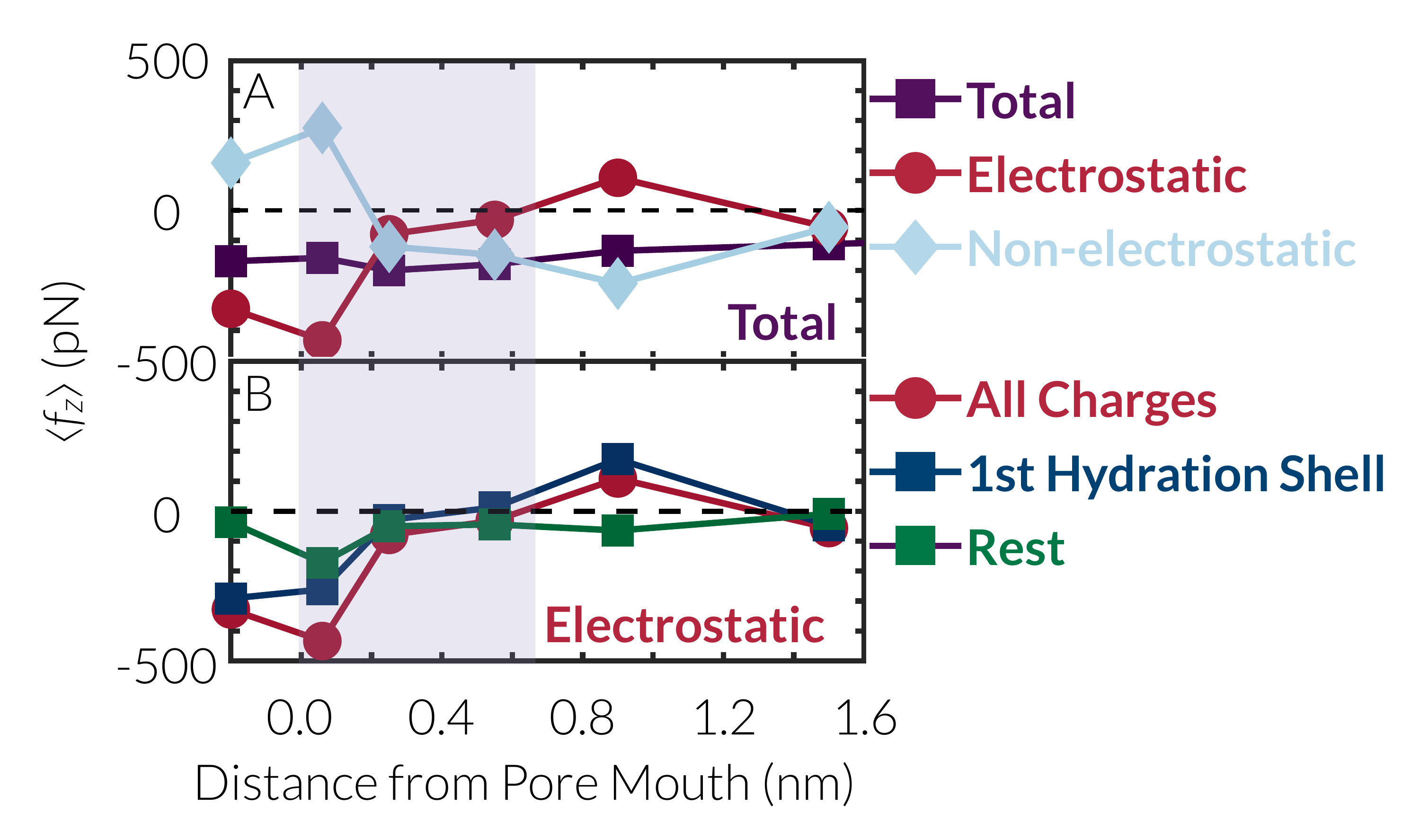}
\caption{\label{fig:elec-cont-Cl}{(A) Electrostatic and non-electrostatic contributions to the total restraining force exerted on the leading chloride at 280~K. (B) The contributions of first hydration shell  and the rest of charges on the electrostatic forces shown in (A). The shaded purple region corresponds to the pore interior.
}}
\end{figure*}

\begin{figure*}[h]
\centering
\includegraphics[width=.81\textwidth]{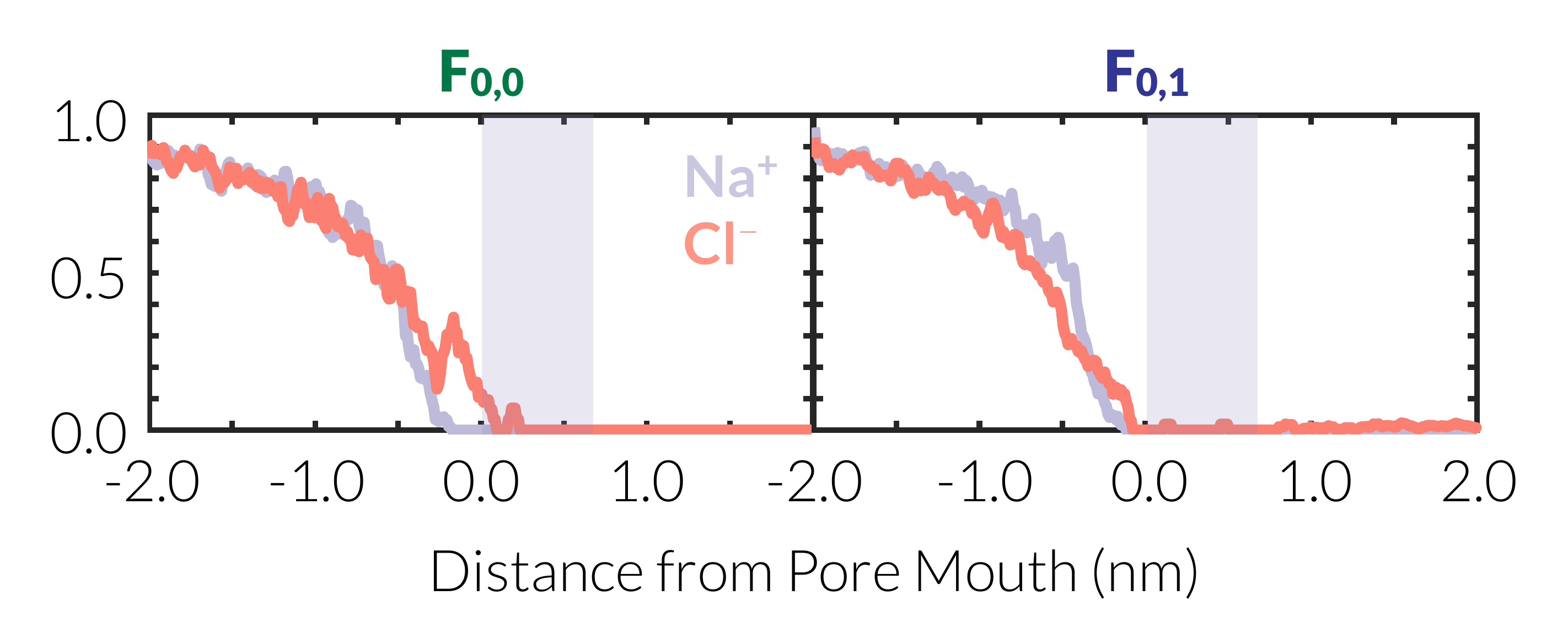}
\caption{\label{fig:ion-dist-basin} Ionic number density as a function of directed curved  distance from pore mouth for sodium (light purple) and chloride (peach) ions at 280~K within the $F_{0,0}$ and $F_{0,1}$ basins. The shaded purple region corresponds to the pore interior.
}
\end{figure*}

\begin{figure*}[h]
\centering
\includegraphics[width=.87\textwidth]{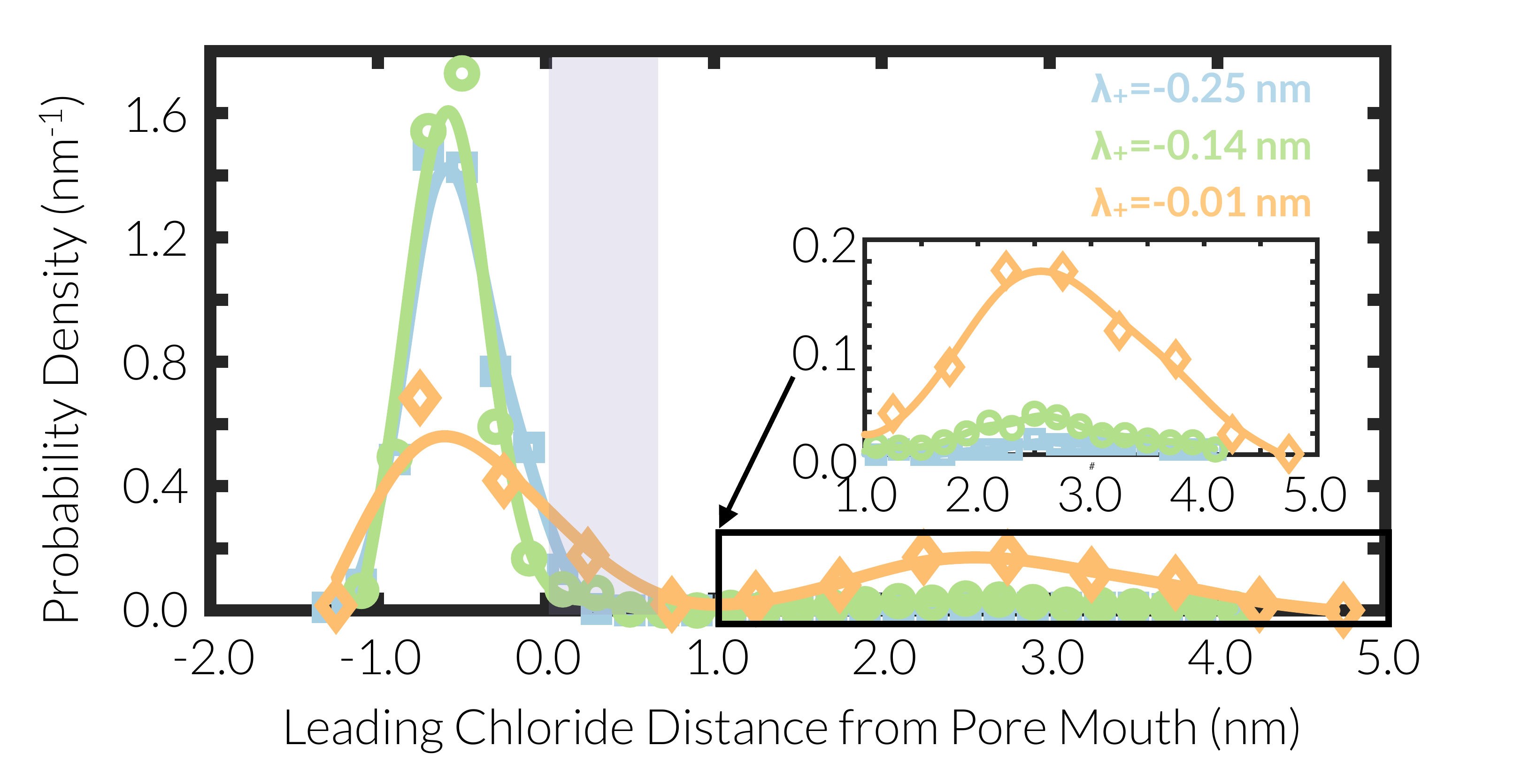}
\caption{{Spatial distribution of the leading chloride in the FFS calculation with $\lambda_+$ at 280~K. The fraction of crossing points with their leading chloride already having traversed the pore increase dramatically from $\lambda_+=-$0.25~nm to $\lambda_+=-$0.01~nm. The shaded purple region corresponds to the pore interior.}}
\end{figure*}

\begin{figure*}[h]
\centering
\includegraphics[width=.88\textwidth]{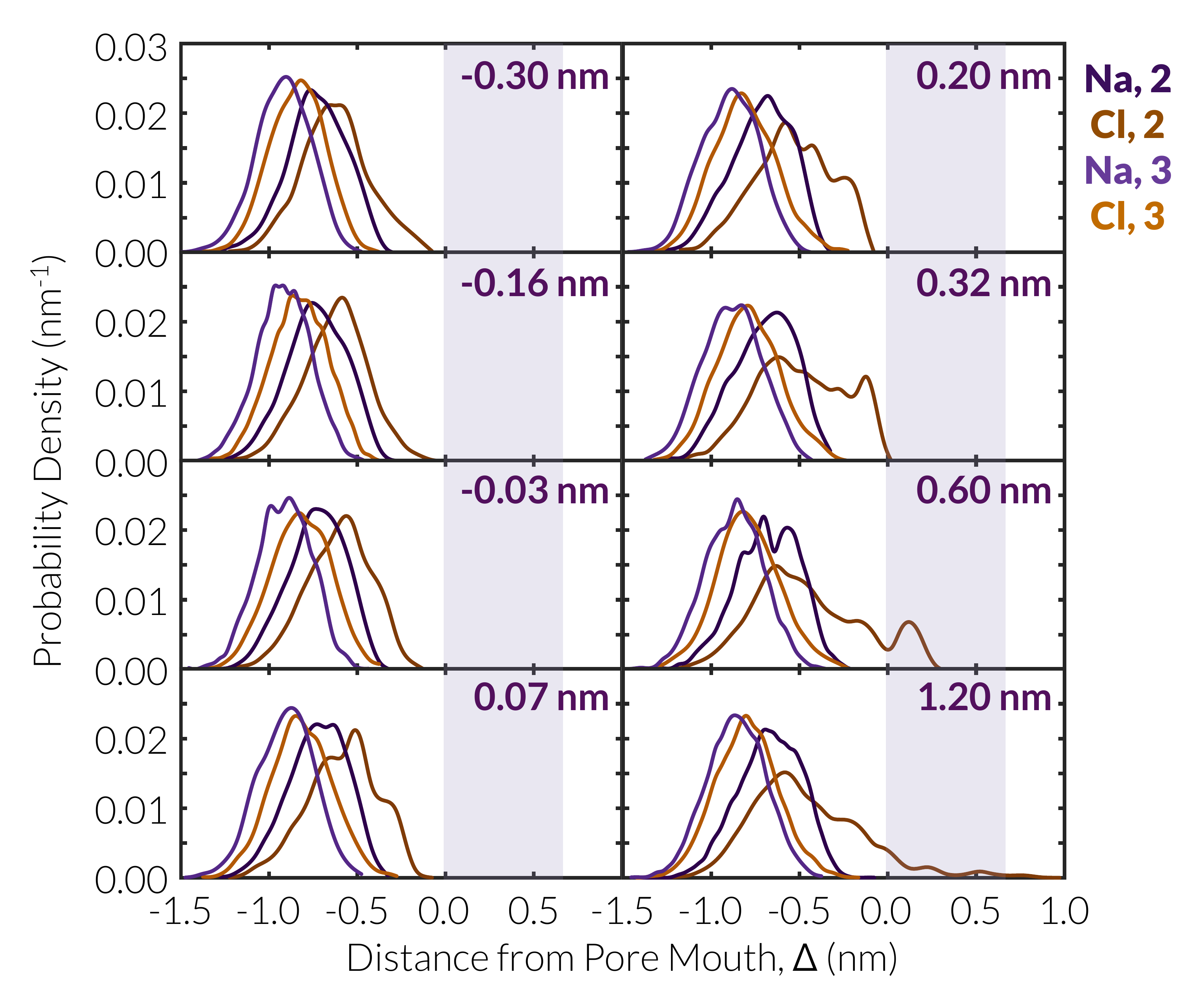}
\caption{{Spatial distribution of non-leading ions during the process of sodium transport at 280~K. The separation between the second chloride and the second sodium becomes fairly pronounced during the transport process. The shaded purple region corresponds to the pore interior.}}
\end{figure*}


\clearpage

\end{document}